\documentclass[aps,prb,showpacs,reprint,superscriptaddress,amsmath,amssymb,nofootinbib]{revtex4-1}

\usepackage[utf8]{inputenc}

\usepackage{gensymb,amsmath,amssymb,graphicx,color}
\usepackage{hyperref}
\usepackage[normalem]{ulem}
\usepackage{multirow}
\usepackage{tikz}

\newcommand{\refcite}[1]{Ref.\,\onlinecite{#1}}

\newcommand{\eqnref}[1]{Eq.\,(\ref{#1})}

\newcommand{\figref}[1]{Fig.\,\ref{#1}}
\newcommand{\sfigref}[2]{Fig.\,\hyperref[#1]{\ref{#1}(#2)}}

\newcommand{\secref}[1]{Sec.\,\ref{#1}}
\newcommand{\appref}[1]{Appendix\,\ref{#1}}

\definecolor{orange}{RGB}{255,165,0}

\definecolor{kspink}{RGB}{200,0,200}

\newcommand{\opics}[2]{$\mathord{\vcenter{\hbox{\includegraphics[scale=#2]{#1}}}}$}
\newcommand{\opic}[1]{\opics{#1}{.8}}

\newtheorem{mydef}{Definition}

\hypersetup{
  colorlinks = true,
  urlcolor = blue,
  pdfauthor = {Wilbur Shirley, Kevin Slagle, and Xie Chen},
  pdftitle = {Twisted foliated fracton phases}
}

\begin{document}

\title{Twisted foliated fracton phases}
\date{\today}

\author{Wilbur Shirley}
\affiliation{Department of Physics and Institute for Quantum Information and Matter, California Institute of Technology, Pasadena, California 91125, USA}
\author{Kevin Slagle}
\affiliation{Department of Physics and Institute for Quantum Information and Matter, California Institute of Technology, Pasadena, California 91125, USA}
\affiliation{Walter Burke Institute for Theoretical Physics, \\
California Institute of Technology, Pasadena, California 91125, USA}
\author{Xie Chen}
\affiliation{Department of Physics and Institute for Quantum Information and Matter, California Institute of Technology, Pasadena, California 91125, USA}
\affiliation{Walter Burke Institute for Theoretical Physics, \\
California Institute of Technology, Pasadena, California 91125, USA}

\begin{abstract}
In the study of three-dimensional gapped models, two-dimensional gapped states should be considered as a free resource. This is the basic idea underlying the notion of `foliated fracton order' proposed in Ref.~\onlinecite{3manifolds}. We have found that many of the known type I fracton models, although they appear very different, have the same foliated fracton order, known as `X-cube' order. In this paper, we identify three-dimensional fracton models with different kinds of foliated fracton order. Whereas the X-cube order corresponds to the gauge theory of a simple paramagnet with subsystem planar symmetry, the novel orders correspond to twisted versions of the gauge theory for which the system prior to gauging has nontrivial order protected by the planar subsystem symmetry. We present constructions of the twisted models and demonstrate that they possess nontrivial order by studying their fractional excitation contents.
\end{abstract}

\maketitle


\section{Introduction}
\label{sec:intro}

The discovery of various ``fracton" models
  \cite{FractonRev,VijayFracton,Sagar16,VijayCL,ChamonModel,HaahCode,YoshidaFractal,Slagle17Lattices,CageNet,VijayNonabelian,MaLayers,ChamonModel2,HsiehPartons,HalaszSpinChains,Slagle2spin,BulmashFractal,BulmashGauging,GaugingPermutation,Haah3manifolds,GeneralizedHaah,YouLitinskiOppen}
  has greatly expanded our understanding of gapped phases in three-dimensional systems. A salient feature characterizing this set of models is the existence of gapped fractional point excitations with restricted mobility. Gapped fracton models\footnote{There are also gapless $U(1)$ fracton models which will not be addressed in this work (see Refs.~\onlinecite{PretkoU1,electromagnetismPretko,Rasmussen2016,PretkoDuality,PretkoGravity,BulmashHiggs,MaHiggs,GromovMultipole,Kumar_Potter_2018,U1FractonGauge,YanPyrochlore,WilliamsonBiCheng,SlagleCurvedU1}).}
  are divided into two major classes according to how the motion of point excitations is constrained: type I and type II. In type II models, fractional point excitations can only move in coordination as a set and individually they cannot move at all. These excitations are said to be immobile and are called `fractons'. In type I models, on the other hand, apart from fracton excitations, there can also exist lineons and planons -- fractional excitations which can move by themselves within a plane or along a line. The restricted mobility of the point excitations leads to various new features in the fracton models: a slow thermalization process,\cite{Bravyi11,PremGlassy,LocalizationFractonicCircuits,fractonScars,fragmentationPollmann} stable extensive ground state degeneracy, unusual entanglement scaling, \cite{FractonEntanglement,ShiEntropy,HermeleEntropy,BernevigEntropy} etc.\cite{YanHolography,BulmashBoundaries,FractonFusion,BrownWilliamsonParallelized,SpuriousEntanglement,YouDefects,KubicaYoshidaUngauging}

\begin{figure}[htbp]
    \centering
    \includegraphics[width=0.48\textwidth]{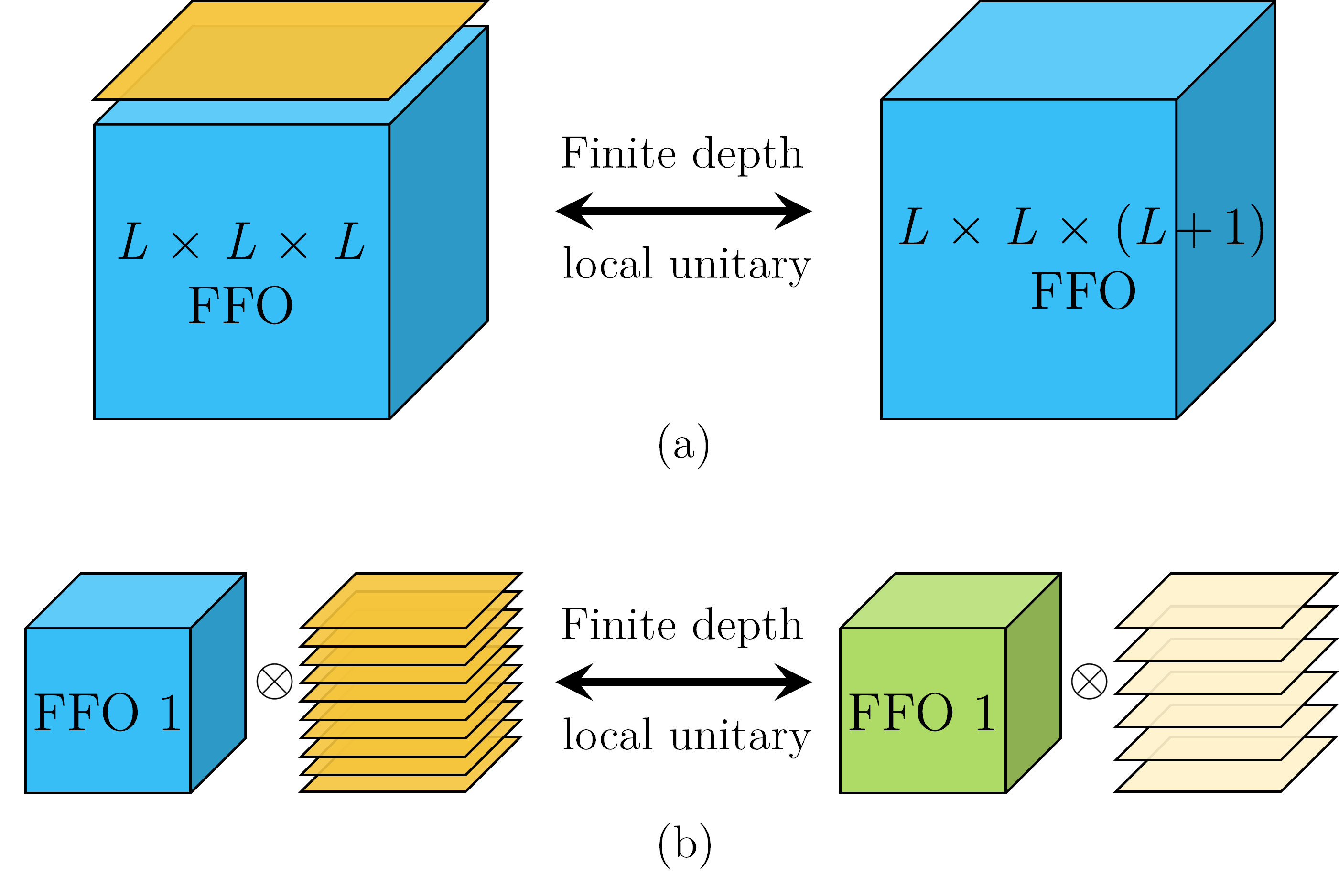}
    \caption{Foliated fracton order: (a) In a model with FFO, different system sizes are related through the addition or removal of 2D layers and finite depth local unitary transformations. (b) Two models have the same FFO if they are related through the addition of decoupled stacks of 2D layers and finite depth local unitary transformation.}
    \label{fig:FFO}
\end{figure}

Among the type I fracton models, we have found that many of them have a hidden `foliation' structure and are said to have `foliated fracton order' (FFO).\cite{3manifolds,FractonEntanglement} That is, starting from a model with a larger system size, we can apply a finite depth local unitary transformation and map the model to a smaller system size together with decoupled layers of 2D gapped states, as illustrated in \sfigref{fig:FFO}{a}. As there should be no fundamental change in the order of the system simply due to the change in system size, we should think of the 2D gapped states as free resources in the study of these 3D fracton models even though the 2D gapped states can have highly nontrivial topological order of their own. Correspondingly, we define two foliated fracton models to have the same `foliated fracton order' if they can be related through a finite depth local unitary transformation upon the addition of decoupled stacks of 2D layers of gapped states, as shown in \sfigref{fig:FFO}{b}.

Using this definition, we have found that many of the type I fracton models with a foliation structure actually have the same foliated fracton order. In particular, we have shown explicitly that the spin checkerboard model, the Majorana checkerboard model, and the semionic X-cube model all have the same FFO as the X-cube model (or multiple copies of it). \cite{Checkerboard,MajoranaCheckerboard,FractonStatistics} The untwisted string-membrane-net model discussed in Ref.~\onlinecite{SlagleSMN} was also shown to be equivalent to the X-cube model. As the X-cube model\cite{Sagar16} can be obtained by gauging the intersecting planar subsystem symmetries of a trivial 3D paramagnet,\cite{Sagar16,GaugingSubsystem} the X-cube FFO is considered to be untwisted. It is similar to the toric code model as an untwisted $Z_2$ gauge theory which can be obtained by gauging the global $Z_2$ symmetry of a trivial 2D paramagnet. It is known that 2D $Z_2$ gauge theory can also be `twisted' where the gauge flux becomes a semionic excitation. It can be obtained from gauging the 2D symmetry protected topological order with $Z_2$ symmetry as shown in Ref.~\onlinecite{LevinGuSPT}. It is then natural to ask whether there exists twisted FFO.

In this paper, we identify three-dimensional fracton models with a `twisted' FFO. That is, these models have an FFO that is different from that of the X-cube model. Moreover, they can be obtained by gauging a 3D model with subsystem planar symmetries that is not a trivial paramagnet. In other words, the ungauged model has (strong) symmetry protected topological order with subsystem planar symmetries. Note that although twisted fracton models have already appeared in the literature, \cite{YouSondhiTwisted,SongTwistedFracton,FractalSPT} they have not been studied in terms of their foliated fracton order. We discuss two (sets of) examples in detail. One is 3-foliated, meaning that we can decouple 2D topological layers in three different directions using finite depth local unitary transformations. The X-cube model is also 3-foliated in this sense and we can consider this new model as the twisted version of two copies of the X-cube model. The other example is 1-foliated, meaning that we can only decouple 2D topological layers in one direction from the model. The untwisted version of a 1-foliated model with $Z_2$ symmetries would simply be a decoupled stack of 2D toric codes.

The paper is organized as follows: In section \ref{sec:3foliated}, we discuss the 3-foliated model by presenting the construction of the model, demonstrating its foliation structure and then showing that its FFO is different from that of the X-cube model. In section \ref{sec:1foliated}, we do the same for the 1-foliated model. We discuss in section \ref{sec:SSPT} how to `ungauge' the models into models with subsystem symmetry protected topological order before summarizing in section \ref{sec:summary}.

\section{Twisted 3-Foliated Model}
\label{sec:3foliated}

In this section, we describe a model that is foliated in the $x$, $y$, and $z$ directions by layers of a twisted 2D $Z_2\times Z_2$ gauge theory. We will see that its foliated fracton order (FFO) is twisted in the sense that its FFO is distinct from that of the X-cube model or copies of it. (A brief review of the X-cube model is given in Appendix \ref{app:Xc}.) Ungauging this fracton model results in a paramagnetic model with (strong) subsystem symmetry protected topological (SSPT) order under 3 sets of intersecting planar subsystem symmetries.

The model is constructed by strongly coupling intersecting layers of a set of 3 perpendicular stacks of twisted 2D $Z_2\times Z_2$ gauge theories, in a manner akin to the construction of the X-cube and semionic X-cube models from stacks of 2D toric code layers and 2D double semion layers respectively. These constructions are discussed in Refs.~\onlinecite{MaLayers} and~\onlinecite{VijayCL}. Like the semionic X-cube model, the 3-foliated model constructed in this section belongs to the class of exactly solvable \textit{twisted fracton models} considered in Ref.~\onlinecite{SongTwistedFracton}. Here, we are able to extend our understanding by studying the model through the lens of the coupled layer construction and as an FFO. Unlike the semionic X-cube model, this $Z_2\times Z_2$ model has twisted FFO; thus, there is a distinction between a fracton model being \textit{twisted} in the sense of Ref.~\onlinecite{SongTwistedFracton}, and a model having \textit{twisted} FFO.

\subsection{Model Construction}
\label{sec:3foliated_model}

\subsubsection{2D $Z_2\times Z_2$ twisted gauge theory}

\label{sec:2D}

First, we briefly review the properties of $Z_2\times Z_2$ twisted gauge theories in 2D, and describe an exactly solvable model for one such theory. Twisted gauge theories may be thought of as Hamiltonian realizations of 2+1d Dijkgraaf-Witten models,\cite{DijkgraafWitten} or as the result of gauging global symmetries in paramagnets with non-trivial symmetry-protected topological (SPT) order.\cite{LevinGuSPT} For $Z_2\times Z_2$ symmetry, there are $2^3=8$ distinct SPT phases in 2D, corresponding to the 8 elements of $H^3(Z_2\times Z_2,U(1))$.\cite{Xie13} They are characterized by the topological invariants $N_1$, $N_2$, and $N_{12}$, each of which takes values 0 or 1. Upon gauging, the exchange statistics of the gauge fluxes are given by $i^{N_1}$ and $i^{N_2}$, whereas the braiding statistics between the two fluxes is $i^{N_{12}}$. In all cases the statistics between gauge charge and corresponding gauge flux is $-1$.\cite{WangLevin15}

Here, we will focus on the twisted gauge theory obtained from the SPT phase with $N_{12}=1$ and $N_1=N_2=0$. In this case, the elementary gauge charges $e_A$ and $e_B$ and bosonic gauge fluxes $m_A$ and $m_B$ obey the following fusion rules:
\begin{equation}
    e_A^2=e_B^2=1\hspace{1cm}
    m_A^2=e_B\hspace{1cm}
    m_B^2=e_A.
\end{equation}
Thus, as an intrinsic topological order, this theory is equivalent to the $Z_4$ toric code, with $m_A$ and $m_B$ mapping onto the $\tilde{e}$ and $\tilde{m}$ $Z_4$ anyons of the $Z_4$ toric code, respectively.

There is a convenient isomorphism between the $Z_4$ clock and shift algebra and the two qubit operator algebra, $\tilde{Z}\to X^AS^B$, $\tilde{X}\to X^BCZ^{AB}$, where $\tilde{Z}$ and $\tilde{X}$ are the clock and shift generators of the $Z_4$ operator algebra with $\tilde{Z}\tilde{X} = i \tilde{X}\tilde{Z}$, $Z$ and $X$ are the clock and shift (Pauli) operators of the $Z_2$ algebra, $S$ is the one-qubit phase gate $\text{diag}(1,i)$, $CZ$ is the two-qubit controlled-Z operator $\text{diag}(1,1,1,-1)$, and $A$ and $B$ label the two qubits. Applied to the $Z_4$ toric code degrees of freedom, this mapping naturally allows one to write the $Z_2\times Z_2$ twisted gauge theory as a $Z_2\times Z_2$ string-net model, such that the gauge charges correspond to violations of the plaquette terms.

\begin{figure}
    \centering
    \includegraphics[width=0.4\textwidth]{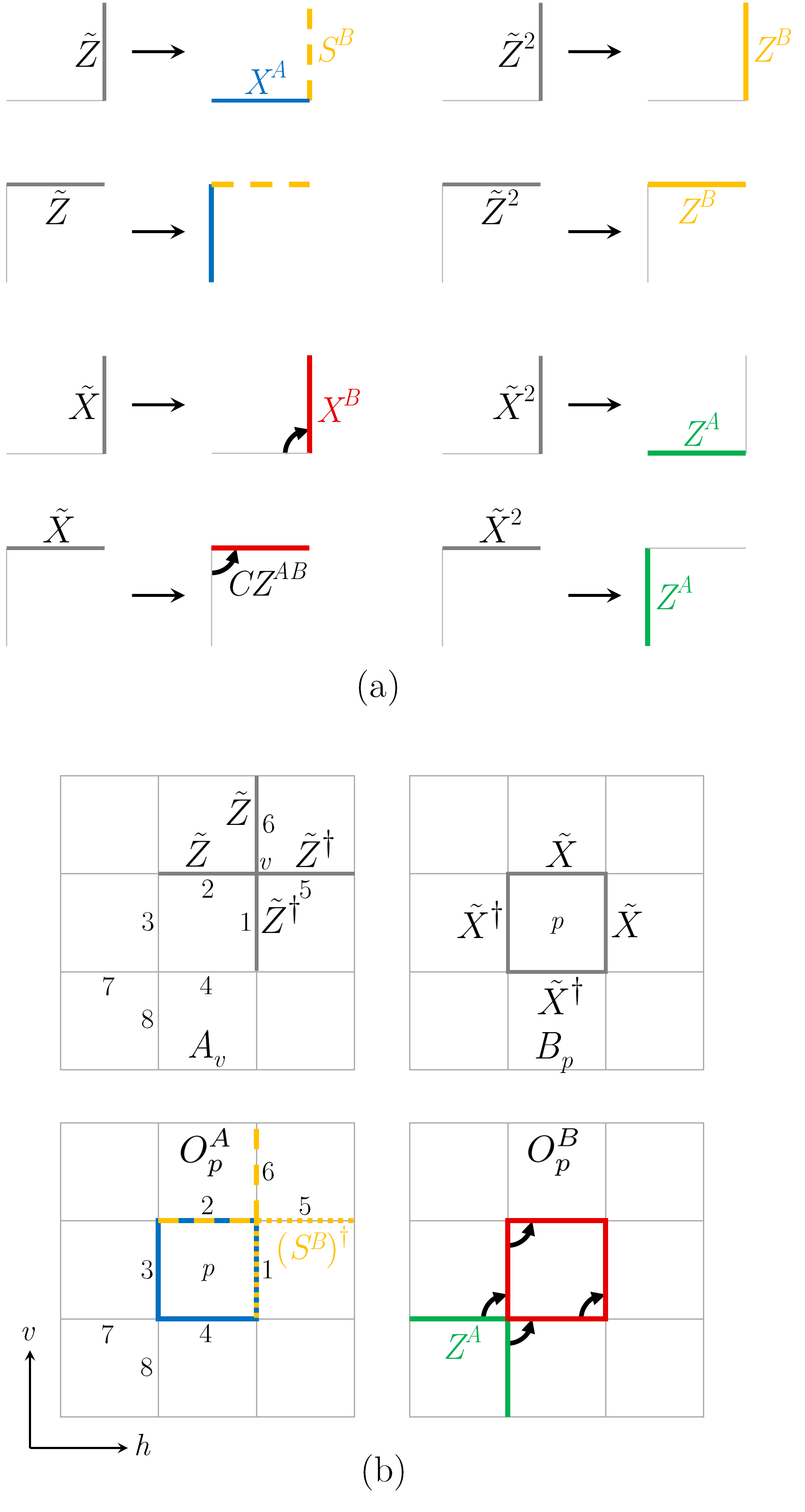}
    \caption{(a) Transformation from $Z_4$ qudit degrees of freedom to two $Z_2$ qubit degrees of freedom. (b) Operators $A_v$, $B_p$, $O_p^A$, and $O_p^B$. In the bottom figures, blue represents $X^A$, red represents $X^B$, dashed yellow represents $S^B$, dotted yellow represents $(S^B)^\dagger$, green represents $Z^A$, and the black arrows represent $CZ^{AB}$ from $A$ to $B$. The action of $Z$, $S$, and $CZ$ gates \textit{precede} the action of the $X$ gates.}
    \label{fig:2D}
\end{figure}

In particular, consider the $Z_4$ toric code Hamiltonian on a square lattice:
\begin{equation}
    H_{TC}=-\sum_v \left(A_v+A_v^2+A_v^3\right)-\sum_p\left(B_p+B_p^2+B_p^3\right)
\end{equation}
where $A_v=\tilde{Z}^\dagger_1 \tilde{Z}^\dagger_5 \tilde{Z}_6\tilde{Z}_2$ and $B_p=\tilde{X}_1 \tilde{X}_2\tilde{X}^\dagger_3 \tilde{X}^\dagger_4$ per Fig. \ref{fig:2D}. The $A_v^2$ and $B_p^2$ terms are redundant, but we keep them in the Hamiltonian so that the transformed Hamiltonian has a convenient correspondence with the string-net formulation.\cite{Stringnet}

After mapping to qubit degrees of freedom and shifting qubit $A$ downward and to the left by half a lattice spacing [as shown in \sfigref{fig:2D}{a}], $H_{TC}$ is transformed into the $Z_2\times Z_2$ twisted gauge theory Hamiltonian
\begin{equation}
    H_{2D}=-\sum_v \left(Q^A_v+Q^B_v\right)-\sum_p\left(O^A_p+O^B_p+\textrm{h.c.}\right),
\end{equation}
where $Q^\mu_v=\prod_{l\in v}Z^\mu_l$, and (see Fig. \ref{fig:2D})
\begin{align}
    O_p^A&=X^A_1X^A_2X^A_3X^A_4\left(S^B_1\right)^\dagger\left(S^B_5\right)^\dagger S^B_6S^B_2, \\
    O_p^B&=X^B_1X^B_2X^B_3X^B_4CZ^{AB}_{32}CZ^{AB}_{41}CZ^{AB}_{73}CZ^{AB}_{84}Z^A_7Z^A_8. \nonumber
\end{align}
In particular, $A_v\to O_p^A$, $A_v^3\to \left(O_p^A\right)^\dagger$, $B_p\to O_p^B$, $B_p^3\to \left(O_p^B\right)^\dagger$, $A_v^2\to Q_v^A$, and $B_p^2\to Q_v^B$.
Note that $\left(X^BCZ^{AB}\right)^\dagger=X^BCZ^{AB}Z^A$.

As this transformation is an exact mapping, it is obviously possible to carry through the following construction in terms of the original $Z_4$ degrees of freedom. As we will see however, the $Z_2\times Z_2$ degrees of freedom provide a more natural language to analyze the emergent fracton order.

\subsubsection{Coupled layers construction}
\label{sec:coupled layers}

\begin{figure}
    \centering
    \vspace{.25cm}
    \includegraphics[width=0.45\textwidth]{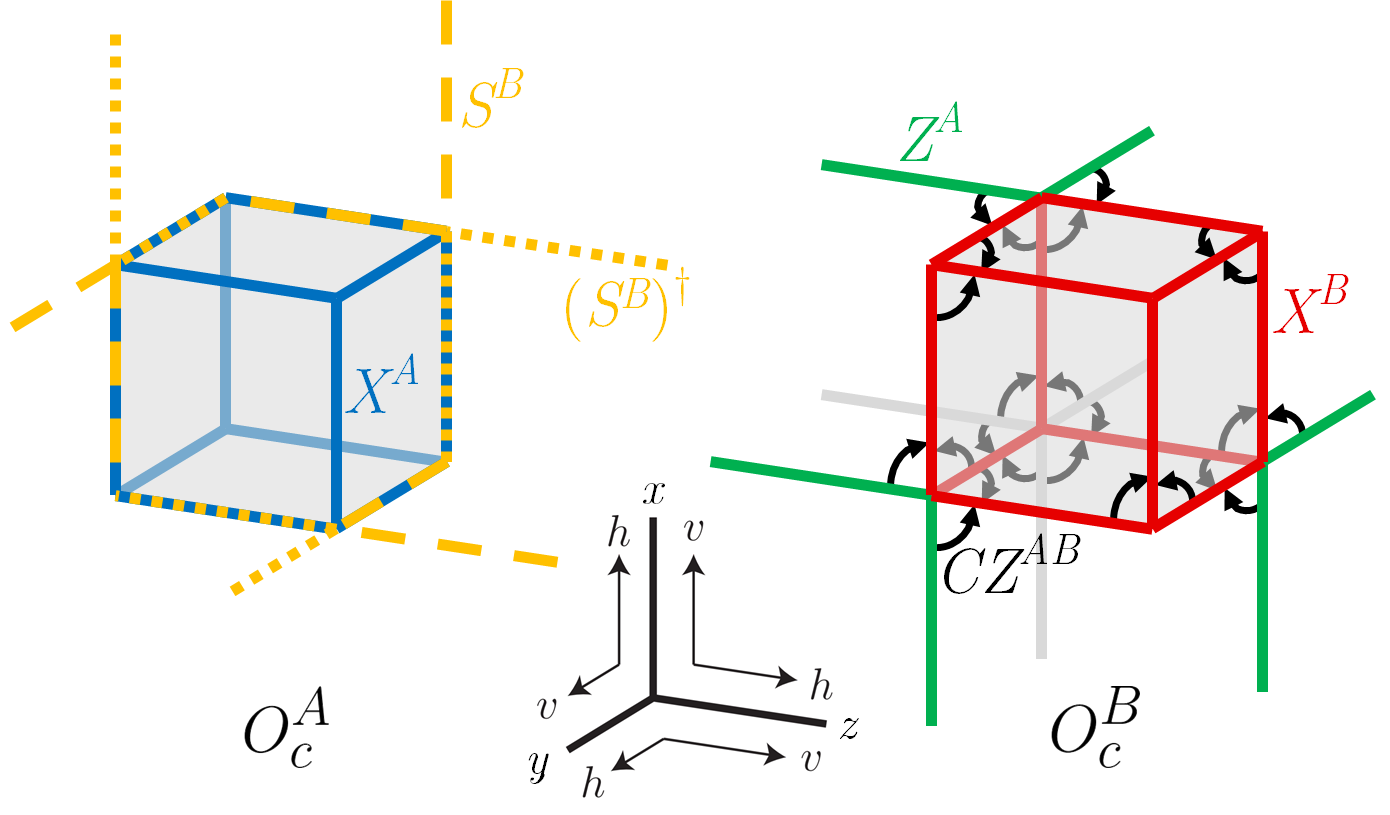}
    \caption{Cube operators of $H_{3D}$. Here, blue represents $X^A$, red represents $X^B$, dashed yellow represents $S^B$, dotted yellow represents $(S^B)^\dagger$, green represents $Z^A$, and the black arrows represent $CZ^{AB}$ from $A$ to $B$. The action of $Z$, $S$, and $CZ$ gates \textit{precede} the action of the $X$ gates.}
    \label{fig:3D}
\end{figure}

The construction of the 3-foliated fracton model is a straightforward generalization of the construction of the X-cube and semionic X-cube models in Refs.~\onlinecite{MaLayers}. We first start with 3 mutually perpendicular intersecting stacks of the $Z_2\times Z_2$ twisted gauge theory model $H_{2D}$, oriented as in Fig. \ref{fig:3D}. Recall that $H_{2D}$ contains 2 qubit degrees of freedom ($A$ and $B$) on each edge of a square lattice. Each edge of the 2D layers coincides with another edge from an orthogonal layer to form a cubic lattice, with each edge containing 4 qubits. Then, couplings of the form $Z^AZ^A$ and $Z^BZ^B$ between qubits on the same edge are added to the Hamiltonian.

In the strong coupling limit, the four qubits at each edge merge into two. The following effective Hamiltonian emerges at lowest order in perturbation theory:
\begin{align}
\begin{split}
    H_{3D}=&-\sum_v\sum_{\sigma=x,y,z} \left(Q^A_{v,\sigma}+Q^B_{v,\sigma}\right) \\
    &-\sum_c\left(O^A_c+O^B_c+\textrm{h.c.}\right),
\end{split}
\label{eq:H3D}
\end{align}
where $v$ runs over vertices of the cubic lattice and $c$ runs over the elementary cubes. $Q^\mu_{v,\sigma}$ are vertex terms equal to products of Pauli $Z^\mu$ operators over the links adjacent to $v$ in the plane normal to $\sigma$. The cube operators $O^A_c$ and $O^B_c$ are depicted graphically in \figref{fig:3D}. The terms of $H_{3D}$ are mutually commuting and unfrustrated and thus the model is exactly solvable. It bears striking similarity to (two copies of) the X-cube model: the vertex terms are identical, and the cube terms are similar in that they involve products of Pauli $X$ operators over the edges of the cube. However, they contain additional phase factors not present in the X-cube terms.

As in the X-cube model, excitations of the vertex constraints are lineons whereas excitations of the cube terms are fractons. Lineons are created at the endpoints of open rigid string operators, whereas fractons are created at the corners of membrane operators. Examples of these operators are given in the discussion of interferometric operators in Sec. \ref{sec:QSS}. Like the X-cube model, planons also exist as fracton dipoles and lineon dipoles, as will be discussed in detail below.

\subsection{Fractional excitations}

In the intermediate coupling regime, the transition to the strong-coupling phase can be thought of as a condensation of $A$ and $B$ type charge loops; correspondingly the ground state of $H_{3D}$ may be viewed as a condensate of charge loops. This mechanism has been studied in detail and dubbed $p$-string condensation in Ref.~\onlinecite{MaLayers}. The structure of excitations in the condensed phase can be understood in terms of the degrees of freedom of the pre-condensed stacks of twisted $Z_2\times Z_2$ gauge theories. Similar to the case of the X-cube model discussed in Refs.~\onlinecite{MaLayers}, the 2D gauge charges of the original decoupled stacks fractionalize into fracton dipoles (a pair of adjacent fractons whose axis is normal to the 2D layer), and remain as $Z_2$ planons. These planons will be labelled $e^A_{\mu\nu,i}$ and $e^B_{\mu\nu,i}$ where $\mu\nu,i$ refers to the plane of mobility ($\mu$ and $\nu$ the planar axes and $i$ the coordinate in the normal direction). In the charge loop picture, individual fractons correspond to endpoints of open charge strings above the condensate. They will be denoted as $f^A_{ijk}$ and $f^B_{ijk}$, where $ijk$ denotes spatial location, and likewise inherit $Z_2$ fusion rules:
\begin{equation}
    \left(f^A_{ijk}\right)^2=\left(f^B_{ijk}\right)^2=\left(e^A_{\mu\nu,i}\right)^2=\left(e^B_{\mu\nu,i}\right)^2=1.
\end{equation}

As in the X-cube coupled layers construction, individual gauge fluxes of the original stacks are confined upon condensation due to their statistical interaction with the charge loops. However, composites of an $A$ ($B$) flux and an $A$ ($B$) anti-flux in orthogonal planes have trivial statistics with the charge loops, and thus survive the condensation. These composites become $A$ and $B$ type lineons of the condensed phase, labelled as $l^A_{\mu,ij}$ and $l^B_{\mu,ij}$ with $\mu$ the axis of mobility and $i$ and $j$ the normal coordinates. By convention $l^A_{\mu,ij}$ ($l^B_{\mu,ij}$) consists of a flux in the $\mu\nu$ plane and an anti-flux in the $\rho\mu$ plane. They inherit the fusion rules from the 2D gauge fluxes, and therefore obey:
\begin{gather}
\begin{split}
    \left(l^A_{\mu,ij}\right)^2=e^B_{\mu\nu,i}\times e^B_{\rho\mu,j}\\
    \left(l^B_{\mu,ij}\right)^2=e^A_{\mu\nu,i}\times e^A_{\rho\mu,j}\\
    \left(l^A_{\mu,ij}\right)^4=\left(l^B_{\mu,ij}\right)^4=1.
    \label{eqn:lineon}
\end{split}
\end{gather}
In these equations, the fracton dipoles' planes of mobility intersect along the lineon axis. There are also triple fusion rules between intersecting lineons along orthogonal axes (coordinate labels have been suppressed):
\begin{equation}
    l^A_{x}\times l^A_{y}\times l^A_{z}=l^B_{x}\times l^B_{y}\times l^B_{z}=1.
\end{equation}

Whereas individual lineons are restricted to move along a line, adjacent lineon anti-lineon pairs, called lineon \textit{dipoles}, are free to move in a plane normal to the axis of separation, and are hence planons. This is because lineons arise as bound states of flux anti-flux pairs in orthogonal planes. A lineon dipole therefore contains four original flux (or anti-flux) excitations. However, the flux anti-flux pair in the plane shared by the two lineons annihilate one another, leaving behind a flux anti-flux pair in adjacent parallel planes. Lineon dipoles will be denoted $m^A_{\mu\nu,i,i+1}$ and $m^B_{\mu\nu,i,i+1}$ where $\mu\nu$ refers to the plane of mobility and $i$ and $i+1$ are the coordinates in the normal direction of the parallel planes containing the flux and anti-flux respectively. The following fusion rules hold by definition:
\begin{align}
\begin{split}
    m^A_{\mu\nu,i,i+1}&=l^A_{\mu,ij}\times \bar{l}^A_{\mu,i+1,j}=l^A_{\nu,ki}\times \bar{l}^A_{\nu,k,i+1}\\
    m^B_{\mu\nu,i,i+1}&=l^B_{\mu,ij}\times \bar{l}^B_{\mu,i+1,j}=l^B_{\nu,ki}\times \bar{l}^B_{\nu,k,i+1},
\end{split}
\label{eqn:ldipoles}
\end{align}
where $\bar{l}$ refers to the anti-lineon of $l$. Combining \eqnref{eqn:lineon} and \eqnref{eqn:ldipoles} yields the rules
\begin{align}
\begin{split}
    \left(m^A_{\mu\nu,i,i+1}\right)^2&=e^B_{\mu\nu,i}\times e^B_{\mu\nu,i+1}\\
    \left(m^B_{\mu\nu,i,i+1}\right)^2&=e^A_{\mu\nu,i}\times e^A_{\mu\nu,i+1}.
\end{split}
\end{align}

The statistics of excitations in the condensed phase can also be inferred from the anyon statistics of the decoupled stacks. In particular, the fracton dipole $e^A_{\mu\nu,i}$ ($e^B_{\mu\nu,i}$) exhibits a $-1$ braiding statistic when wound around type $A$ ($B$) lineons mobile within the dipole's plane of movement. In particular, these lineons are  $l^A_{\nu,ij}$ ($l^B_{\nu,ij}$) and $l^A_{\mu,ji}$ ($l^B_{\mu,ji}$). Moreover, coplanar lineons of opposite species $l^A_{\mu,ij}$ and $l^B_{\nu,ki}$ inherit the $i$ braiding statistic between gauge fluxes $m_A$ and $m_B$; thus they exhibit an $i$ statistical phase upon crossing. This property, along with the lineon fusion rules, are the essential features that distinguish the twisted 3-foliated model from the untwisted version, i.e. two copies of the X-cube model.

\subsection{Foliation structure}

In this section, we first show that the model described in the last section indeed has a foliated fracton order. That is, one can decouple 2D topological layers out of the model while shrinking the system size as shown in Fig.~\ref{fig:FFO} (a). Then we are going to look at some of the universal quantities of foliated fracton orders, including the quotient super-selection sectors and the entanglement signatures that we discussed in Refs.~\onlinecite{FractonStatistics} and \onlinecite{FractonEntanglement}. It turns out that this model is trivial (the same as two copies of the X-cube model) in both aspects. However, it is not equivalent to two copies of the X-cube model as an FFO, which we will show in \secref{sec:twisted order}.

\subsubsection{Resource layers}

In this section, we demonstrate the 3-foliated structure of the model. We show that resource layers consisting of bilayer 2D $Z_2\times Z_2$ twisted gauge theories can be decoupled from the model in all three directions. Rather than finding an exact local unitary transformation, we arrive at this conclusion by examining the structure of fractional excitations in an $L_x\times L_y\times L_z$ size 3D model, and find that it can be decomposed into two parts: one corresponding to a reduced $L_x\times L_y\times (L_z-2)$ size 3D model, and the other corresponding to two layers of the twisted gauge theory described by $H_{2D}$. That is, the superselection sectors of the larger 3D model are identical to those of the smaller 3D model together with the decoupled 2D layers. We may then conclude the presence of such a foliation structure.

In gapped abelian phases, the superselection sectors form an abelian group under fusion. Decomposing this structure therefore amounts to finding a generating set of the fusion group which can be bipartitioned into sets $A$ and $B$ such that there are no statistical interactions between sectors of $A$ and sectors of $B$.

For the model in question, $S$ contains fractons, lineons, and planons. However, the elementary planons are either fracton dipoles or lineon dipoles (lineon anti-lineon pairs). Therefore, fusion with the appropriate planon effectively \textit{transports} lineons or fractons in their directions of immobility. Hence, a generating set of $S$ need only one lineon of each type in each direction, one fracton of each type, and a generating set of the planon subgroup $P\le S$ (i.e. the subgroup of $S$ generated by the set of all planons), which decomposes as $P=P_{xy}\times P_{yz}\times P_{zx}$ for the three different planes of mobility. This phenomenon also occurs in all of the stabilizer code models with FFO that have been previously studied.\cite{} In fact, this observation is the basis of the notion of \textit{quotient superselection sectors} (QSS), which are elements of the quotient group $Q=S/P$ to be discussed below.

Suppose we wish to disentangle a resource layer in the $z$ direction from the twisted 3-foliated model. Due to the above observation, a decomposition $S=S_{2D}\times S'$, where $S_{2D}$ represents a single 2D resource layer and $S'$ is the reduced 3D model, amounts to a decomposition $P=P_{2D}\times P'=\left(P_{2D}\times P_{xy}'\right)\times P_{yz}\times P_{zx}$, such that $P_{2D}$ has no statistical interaction with $P'$. Moreover, $P_{2D}$ must have trivial interactions with the generating lineons and fractons. However, these generators can always be chosen to lie away from the support of the $P_{2D}$ string operators; thus, this latter condition is essentially vacuous.

\begin{figure}
    \centering
    \includegraphics[width=\columnwidth]{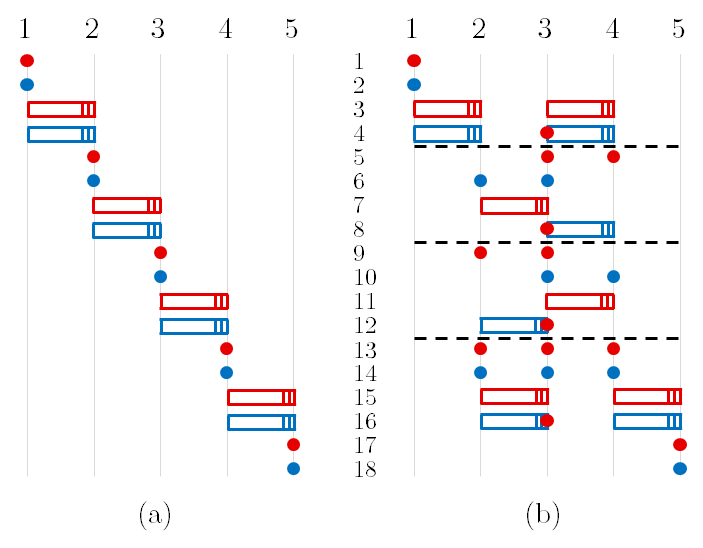}
    \caption{Disentangling an $xy$-plane $Z_2\times Z_2$ twisted gauge theory resource bilayer from the twisted 3-foliated model, in terms of a generating set of the planon excitations. In this notation, the $z$ axis lies along the horizontal direction, with the grid representing the lattice spacing. Each row represents one planon in the generating set. Lineon dipoles $m^A_{xy,i,i+1}$ and $m^B_{xy,i,i+1}$ are represented as respectively blue and red boxes spanning between $z$-coordinates $i$ and $i+1$, with a solid edge to represent the lineon and a triple edge to represent the anti-lineon. Conversely, fracton dipoles $e^A_{xy,i}$ and $e^B_{xy,i}$ are represented as blue and red dots at coordinate $i$. Figure (a) depicts a generating set consisting of all elementary fracton dipoles and lineon dipoles. The generating set of (b) is decomposed into two copies of the $Z_2\times Z_2$ twisted gauge theory between the dashed lines and a reduced generating set for the remaining planons outside the lines, which constitutes a smaller version of the original planon group. Note that there are no non-trivial braiding statistics between the three components.}
    \label{fig:RG}
\end{figure}

Let us now consider $P_{xy}$, the subgroup of $S$ consisting of planons mobile in the $x$ and $y$ directions. A generating set of $P_{xy}$ is given by the set of elementary (minimally separated) fracton and lineon dipoles with $z$-oriented dipolar axis. It is possible to find an equivalent generating set that decouples into two subsets: one generates $P_{xy}'$, a reduced version of $P_{xy}$; the other generates $P_{2D}$, which corresponds to two copies of the 2D $Z_2\times Z_2$ twisted gauge theory modeled by $H_{2D}$. To illustrate this decomposition, it is convenient to use a graphical notation, as shown in Fig. \ref{fig:RG}. In \sfigref{fig:RG}{a}, (part of) the generating set of elementary dipoles is depicted. \sfigref{fig:RG}{b} contains an equivalent but different generating set. In this set, the quasiparticles represented by rows 5-12 are completely decoupled from the remaining planons, in the sense that they form a closed group under fusion and have trivial braiding statistics with the other planons. These quasiparticles represent a generating set of the anyon sectors of two copies of the $Z_2\times Z_2$ twisted gauge theory, i.e. a bilayer (rows 5-8 and rows 9-12). The remaining planons constitute a reduced version of the original planon group with two fewer lattice spacings in the $z$ direction.

Importantly, this mapping of generating planons preserves the locality of the excitations in the $z$ direction. In other words, each element of the generating set moves within a finite region in $z$ before and after the mapping. Therefore, we expect that this mapping of excitations can be realized by a finite depth local unitary transformation with support in the vicinity of the decoupled resource bilayer.

Having established the foliation structure in the 3-foliated model, we now ask if it has the same FFO as the X-cube model (or copies of it). As defined in Refs.~\onlinecite{3manifolds} and \onlinecite{FractonEntanglement}, two gapped models have the same foliated fracton order (FFO) if they can be related by a local unitary transformation upon the possible addition of 2D topological order resource states. While this is a rather coarse equivalence relation, previous works have identified the structure of QSS and interferometric statistics, as well as the entanglement signatures discussed prior, as universal characteristics of FFO.\cite{FractonEntanglement,FractonStatistics} As we are going to see in section \ref{sec:QSS} and \ref{sec:3foliatedCSS}, based on these properties alone it is plausible that the 3-foliated model has the same FFO as two copies of the X-cube model. However, as we are going to show in section \ref{sec:twisted order}, the 3-foliated model actually has a different FFO from two copies of the X-cube model. The QSS and entanglement signature hence provide an insufficient characterization of the universal properties of a foliated fracton phase.

\subsubsection{Quotient superselection sectors and interferometric statistics}
\label{sec:QSS}

Consider the QSS fusion group $Q=S/P$. To reiterate, the essential idea behind QSS is that by modding out the planon subgroup $P$, we obtain a \textit{finite} group which is characteristic of the foliated fracton order of a given model. Since lineon and fracton dipole sectors belong to $P$ for the twisted 3-foliated model, it follows that all lineon superselection sectors $l^A_{\mu,ij}$ ($l^B_{\mu,ij}$) belong to one quotient sector, denoted $l^A_{\mu}$ ($l^B_{\mu}$). Moreover all fracton sectors $f^A_{ijk}$ ($f^B_{ijk}$) belong to a single quotient sector, denoted $f^A$ ($f^B$). These quotient sectors generate the entire group $Q$. 

However, lineon and fracton quotient sectors also obey some relations. First, since $e^A_{\mu\nu,i}\times e^A_{\rho\mu,j}$ and $e^B_{\mu\nu,i}\times e^B_{\rho\mu,j}$ belong to $P$, the lineon fusion rules \eqref{eqn:lineon} imply that $\left(l^A_\mu\right)^2=\left(l^B_\mu\right)^2=1$ as quotient sectors. In other words, the lineon quotient sectors obey $Z_2$ fusion rules while the lineon superselection sectors obey $Z_4$ fusion rules. Second, the lineon triple fusion rules are inherited by the quotient group as
\begin{gather*}
    l^A_x\times l^A_y\times l^A_z=
    l^B_x\times l^B_y\times l^B_z=1.
\end{gather*}
Finally, the fractons sectors obey $\left(f^A\right)^2=\left(f^B\right)^2=1$. Therefore, altogether $Q\cong\left(Z_2\right)^{\times6}$, with the generators $f^A$, $f^B$, $l^A_x$, $l^B_x$, $l^A_y$, and $l^B_y$. This QSS structure is isomorphic (in terms of fusion and particle mobility) to that of two copies of the X-cube model, one corresponding to each of the $A$ and $B$ sectors of $Q$. Recall that the X-cube model has QSS group $\left(Z_2\right)^{\times3}$ with generators $f$, $l_x$, and $l_y$, and triple fusion rule $l_x\times l_y\times l_z=1$.

Interferometric operators for foliated orders, as introduced in Ref.~\onlinecite{FractonStatistics}, are unitary operators with support outside the region $R$, where a point excitation is located, that yield nontrivial statistical phases when acting on excitations belonging to nontrivial elements of $Q$, but act as the identity on excitations in $P$. As discussed in Ref.~\onlinecite{FractonStatistics}, for the X-cube model, there are 8 classes of such operators, which have a $Z_2\times Z_2\times Z_2$ group structure. They include a wireframe operator $W$ which yields a $-1$ phase on the quotient sector $f$, and cylindrical membrane operators $M_x$, $M_y$, and $M_z$. The operator $M_x$ yields a $-1$ phase on the $l_y$ and $l_z$ sectors, and similarily for $M_y$ and $M_z$.

In the twisted 3-foliated model, the structure of interferometric operators is identical to that of two copies of the X-cube model, in terms of the geometry of the operators and their statistical interactions with the QSS. In particular, there are operators $W^A$, $W^B$, $M_x^A$, $M_x^B$, $M_y^A$, $M_y^B$, $M_z^A$, and $M_z^B$. The microscopic form of these operators may be computed by taking products of all the Hamiltonian terms of one kind within a large cubic region: the wireframe operators $W^A$ and $W^B$ correspond to products of cube operators $O^A_c$ and $O^B_c$, whereas the membrane operators correspond to products of the vertex terms. Thus, the membrane operators are simply products of Pauli $Z^A$ or $Z^B$ operators over the support of the membrane, as in (two copies of) the X-cube model, whereas the wireframe operators are more complicated.

The rigid string and membrane operators, which create and transport lineons and fractons, have the identical form as these interferometric operators away from the excitations. The statistical interactions between interferometric operators and QSS can be verified by considering the commutation relations of these microscopic operators. One may also view the interferometric operators as planon loop operators for lineon or fracton dipoles with a macroscopic dipolar length.

\subsubsection{Ground state degeneracy and entanglement signatures}
\label{sec:3foliatedCSS}

\begin{figure*}[htbp]
    \centering
    \begin{tabular}{ccccc}
      $A_e^\text{(cage)}=$ & $B_p^\text{(cage)}=$ & $C_e^\text{(cage)}=$ & $D_p^\text{(cage)}=$ & $E_v^\text{(cage)}=$ \\
        $\tilde{Z}^2\tilde{Z}^2\tilde{Z}^2\tilde{Z}^2$
      & $\tilde{X}^2\tilde{X}^2\tilde{X}^2\tilde{X}^2$
      & $\tilde{Z}^2\tilde{Z}^2$
      & $\tilde{X}^2\tilde{X}^2$
      & $\left(\tilde{X} \tilde{X} \tilde{X}^\dagger \tilde{X}^\dagger\right)^6$ \\
      \opic{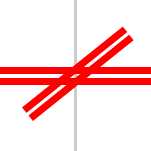} & \opic{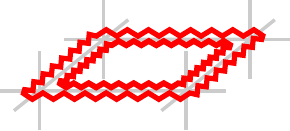} & \opics{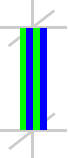}{1} & \opic{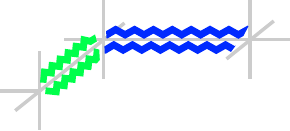} & \opic{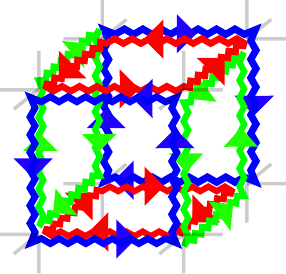} \\
      \opic{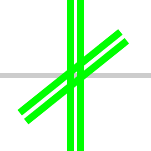} & \opic{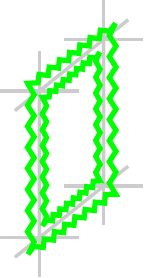} & \opics{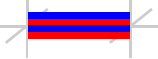}{1} & \opic{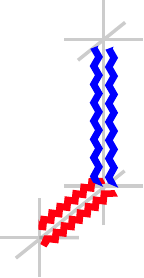}
      & \begin{tabular}{c} \\ \\ \\ \\ $F_v^\text{(cage)}=$ \\
        $\left(\tilde{Z} \tilde{Z} \tilde{Z}^\dagger \tilde{Z}^\dagger\right)^6$ \end{tabular} \\
      \opic{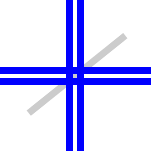} & \opic{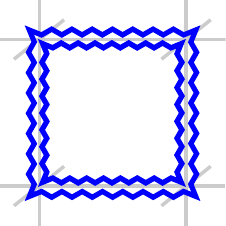} & \opics{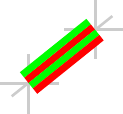}{1} & \opic{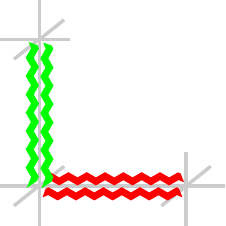} & \opic{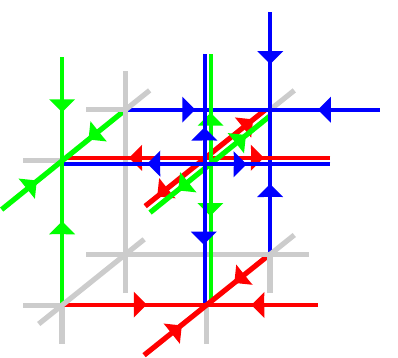}
    \end{tabular}
    \caption{
    A graphical depiction of the terms in the CSS stabilizer version of the 3-foliated model in \eqnref{eq:H3D}.
    Each picture above denotes a term in the stabilizer Hamiltonian.
    There are two $Z_4$ qudits on each edge,
      which will be denoted by two (out of three) different colors.
    Straight colored lines denote $\tilde{Z}$ clock operators,
      while zig-zag colored lines denote $\tilde{X}$ shift operators.
    A conjugate-transpose is taken for operators on edges with arrows that point in the negative x, y, or z direction.
    Double lines denote a $\tilde{Z}^2$ or $\tilde{X}^2$ operator.
    Above each column of pictures, we write the product of operators involved.
    }
    \label{fig:CSS cagenet}
\end{figure*}

To efficiently calculate the entanglement properties of the 3-foliated model, we consider a $Z_4$ CSS stabilizer code formulation of the model.
That is, the Hamiltonian can be expressed as a sum of products of either $\tilde{Z}$ (the clock operator) or $\tilde{X}$ (the shift operator) where all terms in the Hamiltonian commute with each other and each term has eigenvalue $-1$ in the ground state.
This form of Hamiltonian is useful for doing computations, and will allow us to efficiently calculate ground state degeneracy and entanglement entropy.
In \appref{sec:SMN}, we will also express this model in the string-membrane-net and foliated field theory formulations.

To obtain a CSS version of the model, we can repeat the coupled layer construction from \secref{sec:coupled layers},
  but continue using the $Z_4$ clock and shift operators instead of mapping to pairs of qubits. 
The coupled layer construction was performed by adding Pauli $Z^A Z^A$ and $Z^B Z^B$ terms to couple the $Z_2 \times Z_2$ twisted gauge theory layers together.
The $Z^A$ and $Z^B$ operators are written in terms of $\tilde{Z}^2$ and $\tilde{X}^2$, as in
  \sfigref{fig:2D}{a}.
Thus, the $Z^A Z^A$ term that couples $Z_2 \times Z_2$ twisted gauge theory layers is mapped back to a $\tilde{Z}^2 \tilde{Z}^2$ term to couple $Z_4$ toric code layers together.
Unmapping the $Z^B Z^B$ term is similar,
  although note that the $\tilde{X}^2$ operator is not on the same edge as the $Z^B$ operator.
Therefore, the $Z^B Z^B$ term is mapped back to a $\tilde{X}^2 \tilde{X}^2$ operator, but where each $\tilde{X}^2$ is on a different link. The strong coupling limit is described by the CSS code Hamiltonian in \figref{fig:CSS cagenet}.

Since the model is a stabilizer code, we can efficiently calculate its ground state degeneracy and entanglement entropy (see \appref{app:GSD} for details).
The ground state degeneracy of an $L_x\times L_y\times L_z$ system with periodic boundary conditions is
\begin{equation}
  \text{GSD} = 2^{4L_x+4L_y+4L_z-6}. \label{eq:GSD}
\end{equation}

\begin{figure}
    \centering
    \begin{minipage}{.35\columnwidth}\center
      \includegraphics[width=\textwidth]{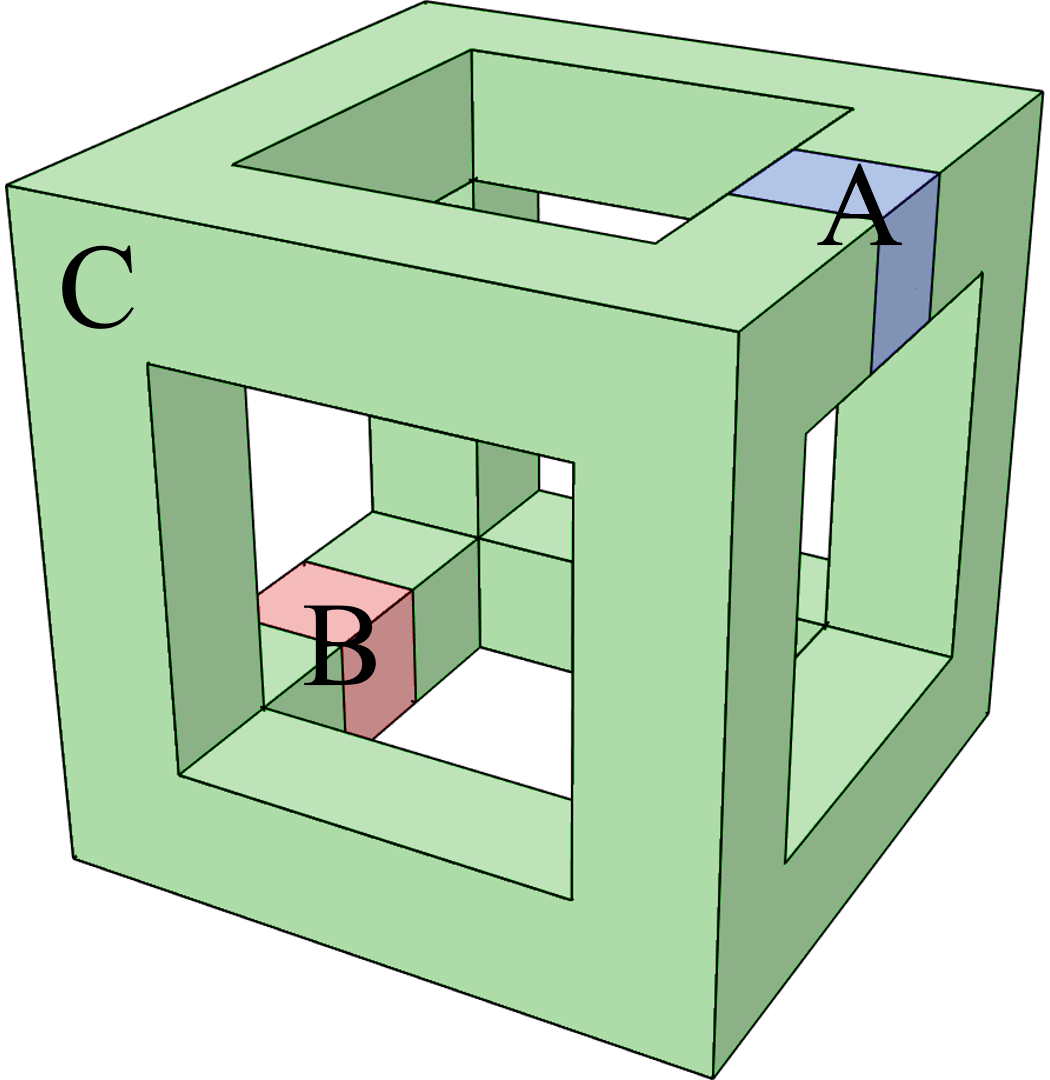}
    \end{minipage}
    \hspace{.6cm}
    \begin{minipage}{.35\columnwidth}\center
      \includegraphics[width=\textwidth]{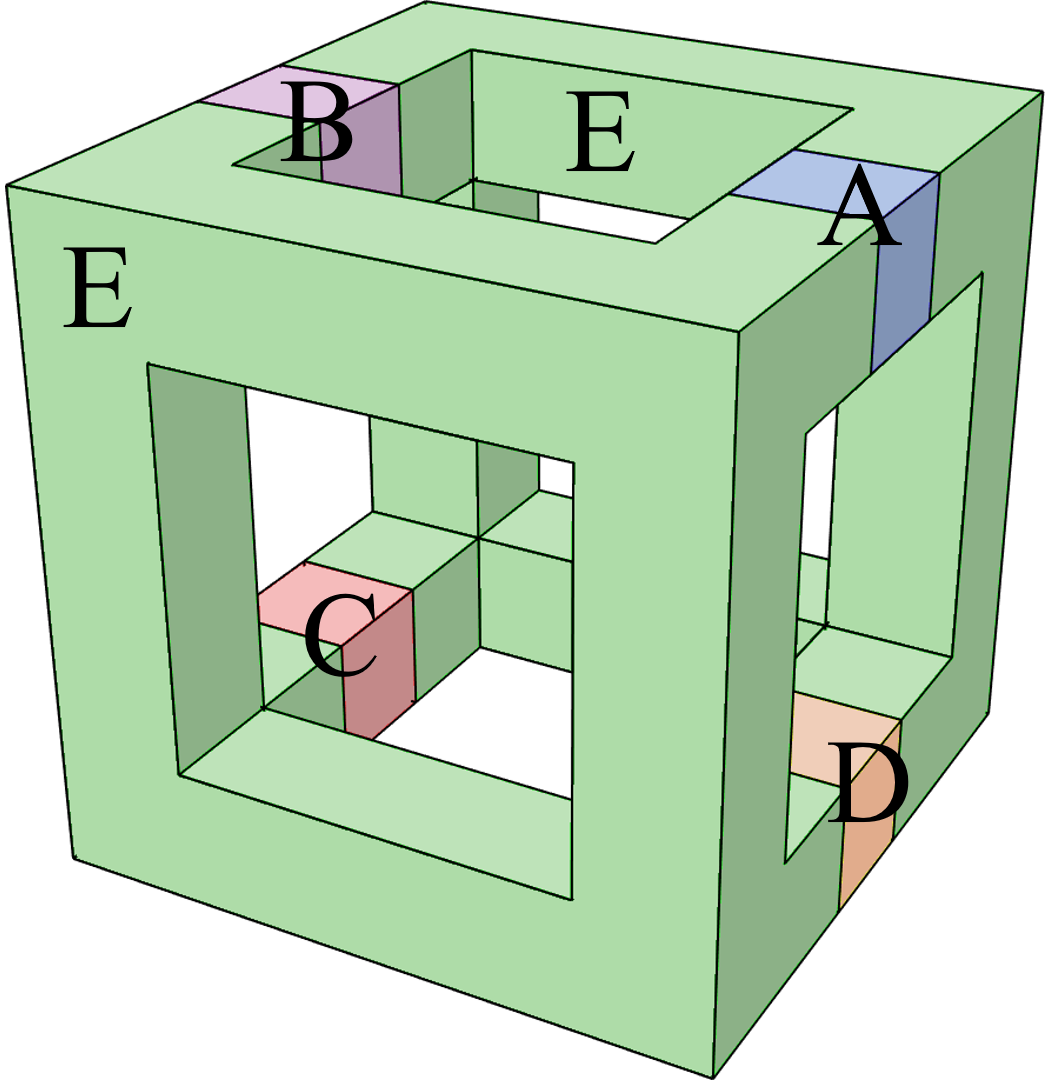}
    \end{minipage} \\ \vspace{.2cm}
    \begin{minipage}{.4\columnwidth}\center
      {\bf (a)} $I(A;B|C)$
    \end{minipage}
    \begin{minipage}{.1\columnwidth}\center
    \end{minipage}
    \begin{minipage}{.4\columnwidth}\center
      {\bf (b)} $I(A;B;C;D|E)$
    \end{minipage}
    \caption{The wireframe geometries used to calculate the entanglement quantities in \eqnref{eq:entanglement}.}
    \label{fig:wireframe}
\end{figure}

Two-dimensional topological orders can be characterized by their topological entanglement entropy. \cite{EntropyKP,EntropyLW}
\refcite{FractonEntanglement} discussed a generalization for foliated fracton orders given by the entanglement quantities $I(A;B|C)$ and $I(A;B;C;D|E)$ computed from subsystems with the wireframe geometries shown in \figref{fig:wireframe}.
For the 3-foliated Hamiltonian (\eqnref{eq:H3D}), we find that
\begin{equation}
  I(A;B|C) = I(A;B;C;D|E) = \log(4). \label{eq:entanglement}
\end{equation}
These entanglement signatures, as well as the ground state degeneracy, are equivalent to that of two copies of the X-cube model.\footnote{%
  In \refcite{FractonEntanglement}, logarithms were evaluated in base 2. With this convention, the entanglement quantities in \eqnref{eq:entanglement} are $I(A;B|C) = I(A;B;C;D|E) = \log_2(4) = 2$. The X-cube model has $I(A;B|C) = I(A;B;C;D|E) = \log_2(2) = 1$.}

\subsection{Twisted foliated fracton order}
\label{sec:twisted order}

While the 3-foliated model appears the same as two copies of the X-cube model in terms of QSS and entanglement signatures, they actually have different FFO. In this section we will demonstrate this difference in two separate ways.

\subsubsection{Lineon fusion rules}

First, we will show that the $Z_4$ fusion rules of the lineon superselection sectors preclude a transformation to two copies of the X-cube model through local unitary and addition of 2D layers. It will be helpful to establish some terminology. A superselection sector that is a fusion product of planons in orthogonal planes, such that the mobility is restricted to the line of intersection of the two planes, will be referred to as a \textit{superficial lineon}. Conversely, a lineon sector that cannot be decomposed as the fusion product of two planons, is referred to as an \textit{intrinsic lineon}.\cite{CageNet}
While intersecting stacks of decoupled 2D topological orders exhibit superifical lineon superselection sectors, only truly fractonic models host intrinsic lineon excitations. 

The key to the argument is that all of the intrinsic lineons in the twisted 3-foliated model are order 4 under fusion (although they square to superficial lineons hence the QSS has order 2), whereas the X-cube model contains intrinsic lineons of order 2. By adding stacks of 2D topological orders, it possible to modify the superselection sector group to include new intrinsic lineons of a \textit{higher} order than the already existing intrinsic lineons. However, the fusion rules of the original intrinsic lineons are immutable, and moreover it is not possible to create a new intrinsic lineon of a \textit{lower} order than the already existing sectors. Therefore, even after the free addition of 2D topological order resource states, the twisted 3-foliated model can never contain intrinsic lineons of order 2. Conversely, the X-cube model, and any number of copies of it, will always retain such a intrinsic lineon. Thus, the two models must have different FFO.

\subsubsection{Redundancies among planons}

Another way to see that the FFO of the 3-foliated model is different from that of two copies of the X-cube model is by looking at the planons. In fact, this can be a useful and generic way to study foliated fracton models. In the following, we are going to show that by examining the planons, we can deduce, first, that the X-cube model is different from a stack of 2D layers and secondly, that the 3-foliated model is different from the X-cube model (or 2 copies of it).

Consider a dimensional reduction procedure from a 3D model to a 2D model where the $x$ and $y$ directions remain infinite while the $z$ direction is made finite. Such a `compactification' process has been used in Ref.~\onlinecite{Arpit_compactify} to study fracton models. We consider the situation where the system has periodic boundary condition in all three directions. As the model is now finite in the $z$ direction, any string operator that extends around the $z$ direction becomes finite and can be added to the Hamiltonian. The ground state degeneracy is reduced and the model becomes a 2D model with anyons moving in the 2D plane. Here, we consider what happens upon this compactification process in three different fracton models: a decoupled stack of 2D layers, the X-cube model and the 3-foliated model. 

We start with a decoupled stack of 2D layers in the $xy$ plane. In the 3D model, there is no string operator in the $z$ direction, therefore after dimensional reduction no extra term can be added. All the planons in the $xy$ planes survive the dimensional reduction. The number of planons grows exponentially with the height of the system in the $z$ direction. We can choose a generating set of all the planons by choosing a generating set for each plane. Such a generating set satisfies the following properties: 
\begin{itemize}
\item Each element in the generating set is constrained to move within a finite segment in $z$ as they come from the 2D layers. We say that the generator planons are `local'. 
\item All other planons that are local can be generated by a subset of the generators that are within a finite distance in $z$. We say that the generating set is `locally complete'. 
\item Moreover, we can make sure the full generating set is not redundant. That is, no element in the generating set (or copies of it) can be generated by other elements in the generating set.
\end{itemize}
For the X-cube model and the 3-foliated model, these properties can no longer be satisfied at the same time.

Now we consider the X-cube model. A brief review of the X-cube model is given in Appendix \ref{app:Xc}. Upon dimensional reduction in the $z$ direction, the string operators in the $z$ direction can be added to the Hamiltonian. Among all the fractional excitations, only the planons in the $xy$ planes survive the dimensional reduction procedure and we can choose a generating set for them consisting of the fracton dipoles $e_i$ centered around plane $i$ and the lineon dipoles $m_{i,i+1}$ living across planes $i$ and $i+1$. Such a generating set is local and locally complete as we defined above. However, it is redundant as the product of all fracton dipoles and the product of all lineon dipoles are both trivial anyons.
\begin{equation}
    \prod_i e_i = 1, \ \prod_i m_{i,i+1}=1
\end{equation}
That is, there exists global constraints among the planons. These global constraints cannot be removed without violating the `locally complete' condition. If we remove $e_1$ and $m_1$ from the generating set, the set is no longer redundant, but $e_1$ and $m_1$ can not be locally generated. Therefore, the X-cube model is different from a stack of 2D layers.

Finally we turn to the 3-foliated model and see how it is different from both the stack of 2D layers and the X-cube model. Upon dimensional reduction, all other superselection sectors are removed except planons in the $xy$ plane, which are the fracton dipoles $e^{A,B}_i$ and lineon dipoles $m^{A,B}_{i,i+1}$. The $e^{A,B}_i$'s and $m^{A,B}_{i,i+1}$ sectors form a locally complete generating set, but it is highly redundant. First, there are local redundancies of the form
\begin{equation}
    \left(m^{A,B}_{i,i+1}\right)^2 = e^{B,A}_{i} \times e^{B,A}_{i+1}
\end{equation}
Moreover, there are global redundancies of the form
\begin{equation}
    \prod_{i} m^{B,A}_{i,i+1} = \prod_{i} e^{A,B}_i = 1
\end{equation}
The global redundancies are similar to that of the X-cube, but the local ones show that the 3-foliated model is different from the X-cube. Note that it is possible to have local redundancy in a locally complete generating set of the X-cube model. For example, if besides all the $e_i$s and and $m_{i,i+1}$s we add $\psi_i = e_i\times m_{i,i+1}$ to the generating set, it will have a local redundancy. However, such local redundancies can be locally removed. That is, if we use the relation $\psi_i = e_i\times m_{i,i+1}$ and eliminate $\psi_i$ from the generating set, we can remove the redundancy. On the other hand, this is not true for the local redundancies in the 3-foliated model. In the 3-foliated model, we can start from the redundancy relation $\left(m^{A,B}_{1,2}\right)^2 = e^{B,A}_1 \times e^{B,A}_2$ and remove it by eliminating $e^{B,A}_2$ from the generating set. Next, we move on to eliminate $e^{B,A}_3$ from the generating set using the redundancy relation $\left(m^{A,B}_{1,2} \times m^{A,B}_{2,3}\right)^2 = e^{B,A}_1 \times e^{B,A}_3$. We can keep doing this, but the redundancy relation that we need to use involves more and more $m$ sectors, and eventually it becomes a non-local relation. We say that the local redundancy relations cannot be locally removed. In fact, a locally complete generating set always has to contain a finite density of $e$ particles and all the $m$ particles, therefore it is always redundant and the redundancy cannot be removed locally. Because of the existence of redundancy relations, especially local redundancy relations that cannot be locally removed, the 3-foliated model is different from both the stack of 2D layers and the X-cube model.

\section{Twisted 1-foliated model}
\label{sec:1foliated}

In this section, we discuss a model which is non-trivially 1-foliated. That is, growing the model in the $z$ direction requires the addition of 2D topological order resource layers ($Z_2\times Z_2$ twisted gauge theories for the model we study), whereas growing the model in the $x$ or $y$ directions simply requires product state resources. At the same time, the model is not local unitarily equivalent to a decoupled stack of 2D topological orders. Nonetheless, all of the fractional excitations of the model are planons, which are mobile in the $xy$ directions; upon compactification in the $z$ direction,\footnote{%
  I.e., a dimensional reduction from a 3D system to a 2D system with a large unit cell.}
the model reduces to a `giant' 2D topological order where the number of superselection sectors grows exponentially with the original height in the $z$ direction.

\subsection{Model construction}

\subsubsection{Boson condensation}

The model is constructed by condensing bosons in a decoupled stack of 2D $Z_2\times Z_2$ twisted gauge theories (equivalently a stack of $Z_4$ toric codes, as discussed in Sec. \ref{sec:2D}), stacked in the $z$ direction. The quasiparticle sectors of the stack consist of $Z_2\times Z_2$ gauge charges $e^A_i$ and $e^B_i$ and gauge fluxes $m^A_i$ and $m^B_i$. Composites of gauge charges in neighboring layers, $e^A_ie^B_{i-1}$, are then condensed to yield a new phase, whose fractional excitations can be understood in the conventional framework of 2D boson condensation in topological phases.\cite{BurnellCondensation,AnyonCondensation}

In particular, charges $e^A_i$ and $e^B_{i-1}$ are identified as a new sector $e_i$. Moreover, individual fluxes are confined due to their non-trivial statistics with the condensed bosons, but flux pairs $m^A_im^B_{i-1}$ survive the condensation as sectors labelled $m_i$. Sectors $e_i$ and $m_i$ have a mutual $-1$ braiding statistic, and adjacent fluxes $m_i$ and $m_{i+1}$ inherit the $i$ braiding statistic. Therefore, the fluxes obey the fusion rules
\begin{equation}
    m_i^2=e_{i-1}\times e_{i+1}.
\end{equation}
Upon compactification, the model may be thought of as a 2D $\prod_{i=1}^L Z_2$ twisted gauge theory with type-II twists \cite{} between adjacent fluxes.

\subsubsection{Giant $K$-matrix}

In 2D, abelian topological orders can be generically understood in terms of the $K$ matrix Chern-Simons formalism.\cite{WenChernSimons} In this description, $N$ species of $U(1)$ gauge fields, $a_I$ with $I=1,\ldots,N$, are governed by the Lagrangian
\begin{equation}
    \mathcal{L}=\frac{1}{4\pi}K_{IJ}\epsilon^{\mu\nu\rho}a^I_\mu\partial_\nu a^J_\rho,
\end{equation}
where $K$ is an $N\times N$ symmetric integer matrix, with even integers along the diagonal for bosonic systems. The quasiparticles are represented by integer vectors $l=(l_1,\ldots,l_N)\in\mathbb{Z}^N$, and have exchange statistics
\begin{equation}
    \theta_l=\pi l^TK^{-1}l,
\end{equation}
whereas their mutual braiding statistics are given by
\begin{equation}
    \theta_{ll'}=2\pi l^TK^{-1}l'.
\end{equation}
Quasiparticles of the form $Kl$ for $l\in\mathbb{Z}^N$ have trivial statistics with all other quasiparticles and thus correspond to local excitations. It is important to note that two $K$ matrices, $K$ and $K'$, are physically equivalent if there is a unimodular matrix $W$ (i.e. with $\det W=1$) such that $K'=W^TKW$. Such a transformation corresponds to a change of quasiparticle basis.

Here, we will employ the $K$ matrix formalism to describe the excitation content of the 3D condensed phase of the prior section. In particular, the structure of planons is captured by a `giant' $N \times N$ $K$-matrix, whose dimension is extensive in the height of the system, and in which spatial locality of excitations in the $z$ direction is encoded in the indices of the vector $l$. In other words, the quasiparticle represented by $l=(\ldots,0,1,0,\ldots)$, with nonzero value at index $I$, is a planon constrained to move near the $xy$ plane with $z$ coordinate equal to $I$ units of the lattice spacing.

We consider the $K$ matrix with the following form in the bulk
(where we have labeled the columns in the anyon basis)
\begin{equation}
\let\quad\; 
\def\-{\raisebox{.75pt}{-}} 
K=\bordermatrix{
     &&e_1&m_1&e_2&m_2&e_3&m_3&e_4&  \cr
&\ddots &  &  &  &  &  &  &  &  \cr
     && 0& 2& \-1&  &  &  &  &  \cr
     && 2& 0&  &  &  &  &  &  \cr
     && \-1&  & 0& 2& \-1&  &  &  \cr
     &&  &  & 2& 0&  &  &  &  \cr
     &&  &  & \-1&  & 0& 2& \-1&  \cr
     &&  &  &  &  & 2& 0&  &  \cr
     &&  &  &  &  &\-1 &  & 0&  \cr
     &&  &  &  &  &  &  &  & \ddots
}. \label{eq:K}
\end{equation}
The inverse matrix $K^{-1}$ has the following form:
\begin{equation}
\let\quad\;
K^{-1}=\frac{1}{4}\bordermatrix{
     && m_0&e_1&m_1&e_2&m_2&e_3&m_3&  \cr
&\ddots &  &  &  &  &  &  &  &  \cr
 && 0 &  & 1 &  &  &  &  &  \cr
 &&  & 0 & 2 &  &  &  &  &  \cr
 && 1 & 2 & 0 &  & 1 &  &  &  \cr
 &&  &  &  & 0 & 2 &  &  &  \cr
 &&  &  & 1 & 2 & 0 &  & 1 &  \cr
 &&  &  &  &  &  & 0 & 2 &  \cr
 &&  &  &  &  & 1 & 2 & 0 &  \cr
 &&  &  &  &  &  &  &  & \ddots 
}. \label{eq:Ki}
\end{equation}
The quasiparticle statistics can be read off from $K^{-1}$. Denoting by $l_I$ the unit vector with all entries equal to 0 except the entry at index $I$, the giant $K$ matrix corresponds precisely to the excitation content of the boson-condensed phase under the assignment $l_{2i-1}=e_i$ and $l_{2i}=m_i$.
In \appref{app:1foliatedLattice}, we describe a lattice model realization of the above $K$-matrix.

\subsection{Foliation structure}

The foliation structure of the model can be easily understood in the $K$-matrix formalism. A single layer of $Z_2\times Z_2$ twisted gauge theory may be disentangled from the bulk via a local unitary transformation represented by the following $W$ matrix, which maps the $e_i$ and $m_i$ anyon basis to a new $\tilde{e}_i$ and $\tilde{m}_i$ basis:
\begin{equation}
\let\quad\;
\def\-{\raisebox{.75pt}{-}}
W=\bordermatrix{
 && \tilde{e}_1&\tilde{m}_1&\tilde{e}^A&\tilde{m}^A&\tilde{e}^B&\tilde{m}^B&\tilde{e}_2&\tilde{m}_2&\tilde{e}_3&   \cr
&\ddots &  &  &  &  &  &  &  &  & &  \cr
e_1&& 1 &  &  &  &  &  &  &  & &  \cr
m_1&&  & 1 &  &  &  & 1 &  &  & &  \cr
e_2&&  &  & 1 &  &  &  & \-1 &  & &  \cr
m_2&&  &  &  & 1 &  &  &  &  & &  \cr
e_3&& \-1 &  &  &  & 1 &  &  &  & &  \cr
m_3&&  &  &  &  &  & 1 &  &  & &  \cr
e_4&&  &  &  &  &  &  & 1 &  & &  \cr
m_4&& \-1 &  &  & 1 &  &  &  & 1 & &  \cr
e_5&&  &  &  &  &  &  &  &  & 1 & \cr
 &&  &  &  &  &  &  &  &  & & \ddots 
}.
\end{equation}
\setcounter{MaxMatrixCols}{20}
This $W$ matrix transforms the $K$-matrix as follows:
\begin{equation*}
\let\quad\;
\def\-{\raisebox{.75pt}{-}}
W^{T}KW=\bordermatrix{
 && \tilde{e}_1&\tilde{m}_1&\tilde{e}^A&\tilde{m}^A&\tilde{e}^B&\tilde{m}^B&\tilde{e}_2&\tilde{m}_2&\tilde{e}_3&  \cr
&\ddots &  &  &  &  &  &  &  &  &  &  \cr
 && 0 & 2 &  &  &  &  & \-1 &  &  &  \cr
 && 2 & 0 &  &  &  &  &  &  &  &  \cr
 &&  &  & 0 & 2 & \-1 & 0 &  &  &  &  \cr
 &&  &  & 2 & 0 & 0 & 0 &  &  &  &  \cr
 &&  &  & \-1 & 0 & 0 & 2 &  &  &  &  \cr
 &&  &  & 0 & 0 & 2 & 0 &  &  &  &  \cr
 && \-1 &  &  &  &  &  & 0 & 2 & \-1 &  \cr
 &&  &  &  &  &  &  & 2 & 0 &  &  \cr
 &&  &  &  &  &  &  & \-1 &  & 0 &  \cr
 &&  &  &  &  &  &  &  &  &  & \ddots
}.
\end{equation*}
Evidently, the transformed $K$-matrix is block diagonal. The $4\times4$ block
(for anyons $\tilde{e}^A$ through $\tilde{m}^B$),
which we will call $K_{2D}$, represents a disentangled copy of 2D $Z_2\times Z_2$ twisted gauge theory. To see that this is the case, note that $K_{2D}$ has inverse
\begin{equation}
\let\quad\;
    K_{2D}^{-1}=\frac{1}{4}\bordermatrix{
    &\tilde{e}^A&\tilde{m}^A&\tilde{e}^B&\tilde{m}^B \cr
    &0 & 2 & 0 & 0 \cr
    &2 & 0 & 0 & 1 \cr
    &0 & 0 & 0 & 2 \cr
    &0 & 1 & 2 & 0 
    }.
\end{equation}
On the other hand, it can easily be seen that the remaining rows and columns represent a smaller version of the original 3D model.

Because the ground state degeneracy of the system only grows with linear system size in the $z$ direction but not in the $x$ and $y$ direction, the model is 1-foliated. That is, growing the model in the $z$ direction requires the addition of 2D topological order resource layers, whereas growing the model in the $x$ or $y$ directions simply requires product state resources.

\subsection{Nontrivial foliated fracton order}

By examining the structure of the planon fusion group, we will demonstrate in this section that the 1-foliated model is not local unitarily equivalent to any decoupled stack of 2D topological orders, nor can it be made equivalent by adding any number of 2D topological order resource layers. In other words, the model represents a non-trivial foliated fracton phase. It is twisted in the sense that ungauging the model yields a nontrivial SSPT phase with 1 set of planar subsystem symmetries.

The situation is very similar to that of the 3-foliated model after dimension reduction. We can choose a locally complete generating set for the planons as $\{e_i,m_i, i =1,...,L\}$. This generating set is redundant with local redundancy relations
\begin{equation}
  m_i^2 = e_{i-1} \times e_{i+1}  
\end{equation}
We can start to remove the redundancy relations by eliminating the $e$'s from the generating set. However, the redundancy relations necessarily gets longer into the form $\left(m_2\times m_4 \times ...\times m_{2n} \right)^2= e_1\times e_{2n+1}$. Therefore, the redundancy relations cannot be locally removed and we conclude that the 1-foliated model is not equivalent to a stack of 2D layers and is hence `twisted'.

\section{Mapping to subsystem SPT phases}
\label{sec:SSPT}

The 3-foliated and 1-foliated model introduced in the previous two sections can be `ungauged' into subsystem symmetry protected topological (SSPT) models \cite{YouSSPT, DevakulSSPT, StephenCellularAutomata}. As the fracton models have twisted foliated fracton order, correspondingly the ungauged model has nontrivial SSPT order. In this section, we first demonstrate how the mapping works, then explain in detail our definition of SSPT order, especially a subtle difference from that given in Ref.~\onlinecite{YouSSPT, DevakulSSPT}.

\subsection{The mapping}

As the 3-foliated model has a `cage-net' type construction\cite{CageNet} as discussed in section~\ref{sec:3foliated_model}, it can be `ungauged' through a duality transformation similar to that described in Ref.~\onlinecite{YouSondhiTwisted} (see also Refs.~\onlinecite{Sagar16,WilliamsonUngauging}). In particular, the `matter' degrees of freedom $\sigma^A$ and $\sigma^B$ live at the center of the cubes in the cubic lattice. The $\sigma$s can be chosen as spin $1/2$ degrees of freedom with on-site symmetry generated by $\sigma^A_x$ and $\sigma^B_x$. Upon `ungauging', the fracton Hamiltonian in Eq.~\ref{eq:H3D} gets mapped to a model of the $\sigma$s with planar subsystem symmetry. The Hamiltonian is
\begin{equation}
    H_{\text{SSPT}} = -\sum_c \left(\tilde{O}^A_c +\tilde{O}^B_c +\text{h.c.}\right)
\end{equation}
where $\tilde{O}^A_c$ and $\tilde{O}^B_c$ are obtained from $O^A_c$ and $O^B_c$ of Eq.~\ref{eq:H3D} in the following way: (1) Replace the tensor product of $12$ $X^A$ ($X^B$) on the edges around the cube $c$ in $O^A_c$ ($O^B_c$) with the matter DOF $\sigma^A_{x,c}$ ($\sigma^B_{x,c}$) at the center of the cube. (2) Replace $Z^A_e$ ($Z^B_e$) on each edge with the tensor product of $4$ $\sigma^A_{z,c}$'s ($\sigma^B_{z,c}$'s) in the cubes containing the edge. Note that the phase factors in the $O^A$ and $O^B$ terms can always be expanded in the basis of $Z^A$ and $Z^B$ operators.\footnote{%
  For example, $S = \frac{1+i}{2} + \frac{1-i}{2}Z$ and $CZ = \frac{1}{2} (1\otimes1+Z\otimes1+1\otimes Z-Z\otimes Z)$.}
Therefore, these replacement steps completely determine the $\tilde{O}$ terms from the $O$ terms. Moreover, as the $\sigma^A_z$ and $\sigma^B_z$ terms always appear as the tensor product of four around each edge, the new Hamiltonian terms are invariant under subsystem planar symmetries 
\begin{equation}
    U^\alpha_{P_{\mu\nu}} = \prod_{c\in P_{\mu\nu}} \sigma^\alpha_{x,c} \;\;\text{ with } \begin{array}{ll} \;\,\alpha=A,B \\ \mu\nu=xy,yz,zx \end{array}
\end{equation}
where $P_{xy}$, $P_{yz}$, $P_{zx}$ denote planes in the $xy$, $yz$, $zx$ direction respectively.

For the 1-foliated model, which is obtained by condensing $e^B_{i-1}e^A_i$ charge pairs in a stack of $Z_2\times Z_2$ twisted gauge theory models, the corresponding SSPT can be obtained from a stack of $Z_2\times Z_2$ twisted SPT \cite{Xie13,WangLevin15} by condensing $e^B_{i-1}e^A_i$ charge pairs. In the SSPT model, condensing charge pairs simply means that the $Z^B_2$ symmetry of the $(i-1)$th layer is combined with $Z^A_2$ symmetry of the $i$th layer into a single symmetry generator. That is, the Hamiltonian of the system is the same as that of a decoupled stack of $Z_2\times Z_2$ twisted SPT, while the planar symmetry generators are tensor products of planar symmetry generators of the $B$ part in layer $i-1$ and the $A$ part in layer $i$.

\subsection{Definition of SSPT order}

As the SSPT models are obtained by `ungauging' twisted fracton models, we expect the SSPT to be `twisted' as well. To be more precise, a 3D system is said to have planar \textit{subsystem symmetry protected topological} (SSPT) order if
\begin{mydef}
The model has a unique symmetric gapped ground state on any closed 3D manifold, which in the absence of symmetry can be mapped to a product state using a finite depth quantum circuit.
\end{mydef}
Two SSPT models with the same subsystem symmetry are said to have the same SSPT order if
\begin{mydef}
The two models can be mapped to each other by adding 2D SPT layers with independent planar symmetries to each model and applying a symmetric finite depth quantum circuit.
\end{mydef}
Note that there is some subtlety in comparing the subsystem symmetry group of two models as the total symmetry group depends on system size. We consider two subsystem symmetry groups to be the same if they can be made the same by adding independent planar symmetry generators to either side.

Accordingly,
\begin{mydef}
An SSPT model has nontrivial or `twisted' SSPT order if it does not have the same SSPT order as a trivial paramagnet (a product state) with the same subsystem symmetry.
\end{mydef}

It is easy to see that once the planar symmetries are gauged, this definition of SSPT order matches the definition of foliated fracton order illustrated in Fig.~\ref{fig:FFO}. This definition can be generalized to models and subsystem symmetries in other dimensions in a straight forward way. 

Our definition is similar but also different from that in Ref.~\onlinecite{YouSSPT, DevakulSSPT}. The definition of Ref.~\onlinecite{YouSSPT, DevakulSSPT} makes use of a `linearly symmetric local unitary circuit' while we use only symmetric finite depth circuits but allow the addition of SPT layers. That is, we require each unitary gate in the circuit to be symmetric while the definition in Ref.~\onlinecite{YouSSPT, DevakulSSPT} allows the individual gates to break symmetry and requires only a subsystem (linear or planar) composite of them to be symmetric. A common consequence of these two definitions is that a pure stack of lower dimensional SPTs, where the subsystem symmetry acts as a global symmetry on each of them, is considered to be a trivial SSPT. On the other hand, the `linearly symmetric local unitary' equivalence is stronger. In particular, in our definition we require the added SPT to come with their own independent symmetry generators. After they are added to the total system, the total subsystem symmetry group is always enlarged. The effect of the `linearly symmetric local unitary' can also be interpreted as allowing the addition of subsystem SPTs. But once added, the symmetry generator of the SPT can be identified with one of the original symmetry generators of the system, hence directly changing the SPT signature associated with that generator. Our definition of equivalence is weaker (e.g. our definition classifies more models as nontrivial) and we have chosen it so that it matches with our definition of foliated fracton order once the subsystem symmetries are gauged.

Upon gauging, the equivalence condition in  Ref.~\onlinecite{YouSSPT, DevakulSSPT} is different from the foliated fracton equivalence we used in this paper. Compared to the foliated fracton equivalence, it amounts to allowing charge condensation in fracton models, because prior to gauging the symmetry group does not necessarily become larger when SPT layers are added. The 1-foliated model discussed above is a trivial SSPT phase under their definition, while it is nontrial under our definition. The 3-foliated model is likely a nontrivial SSPT phase under both definitions.

\section{Summary}
\label{sec:summary}


To summarize, in this paper we demonstrate the existence of twisted foliated fracton order, i.e. 3D gapped fracton models with a foliation structure but which are inequivalent to (copies of) the X-cube model. In particular, we discussed a 3-foliated model in section~\ref{sec:3foliated} and a 1-foliated model in section~\ref{sec:1foliated}. We demonstrated the nontriviality of the models by studying the fractional excitations -- the lineons and the planons -- of the models. In particular, we used a dimensional reduction procedure to reduce the 3D model to a 2D model while keeping track of the locality of the planons along the reduced dimension. By studying the group structure of the local planons, we can discern the differences between stacks of 2D layers, the X-cube model and the twisted models. By using an ungauging procedure, we further mapped the twisted fracton models to nontrivial subsystem symmetry protected topological models.

\section*{Acknowledgments}
We are grateful for helpful discussions with Fiona Burnell, Trithep Devakul, Yizhi You, Hao Song, Abhinav Prem, Xiuqi Ma, Tina Zhang, Roger Mong, Dominic Williamson, and Meng Cheng. W.S. and X.C. are supported by the National Science Foundation under award number DMR-1654340 and the Institute for Quantum Information and Matter at Caltech. X.C. is also supported by the Alfred P. Sloan research fellowship and the Walter Burke Institute for Theoretical Physics at Caltech.
K.S. is supported by the Walter Burke Institute for Theoretical Physics at Caltech.

\appendix

\section{The X-cube model}
\label{app:Xc}

\begin{figure}
    \begin{tikzpicture}
        \pgfmathsetmacro{\l}{1.75}

        \draw[line width=.8,color={rgb:black,1;blue,1}] (0,0,0) -- ++(-\l,0,0) -- ++(0,-\l,0) -- ++(\l,0,0) -- cycle;
        \draw[line width=.8,color={rgb:black,1;blue,1}] (0,0,-\l) -- ++(-\l,0,0) -- ++(0,-\l,0) -- ++(\l,0,0) -- cycle;
        
        \draw[line width=.8,color={rgb:black,1;blue,1}] (0,0,0) -- ++(0,0,-\l*.25);
        \draw[line width=.8,color={rgb:black,1;blue,1}] (0,0,-\l) -- ++(0,0,\l*.25);
        \draw[line width=.8,color={rgb:black,1;blue,1}] (0,-\l,0) -- ++(0,0,-\l*.25);
        \draw[line width=.8,color={rgb:black,1;blue,1}] (0,-\l,-\l) -- ++(0,0,\l*.25);
        \draw[line width=.8,color={rgb:black,1;blue,1}] (-\l,0,0) -- ++(0,0,-\l*.25);
        \draw[line width=.8,color={rgb:black,1;blue,1}] (-\l,0,-\l) -- ++(0,0,\l*.25);
        \draw[line width=.8,color={rgb:black,1;blue,1}] (-\l,-\l,0) -- ++(0,0,-\l*.25);
        \draw[line width=.8,color={rgb:black,1;blue,1}] (-\l,-\l,-\l) -- ++(0,0,\l*.25);
        
        \pgfmathsetmacro{\ri}{1}
        \pgfmathsetmacro{\di}{.3}
        \draw[line width=.5] (.45*\l+\ri,-\l/2-\di,-\l) -- ++(0,0,\l);
        \draw[line width=1,color={rgb:black,.75;red,1}] (\ri+.125,-\l/2-\di,-\l/2) -- ++(\l*.75,0,0);
        \draw[line width=1,color={rgb:black,.75;red,1}] (.45*\l+\ri,-\l*.85-\di,-\l/2) -- (.45*\l+\ri,-\l*.15-\di,-\l/2);

        \pgfmathsetmacro{\rii}{\ri+1.3*\l}
        \draw[line width=.5] (\rii,-\l/2-\di,-\l/2) -- ++(\l*.4,0,0);
        \draw[line width=1,color={rgb:black,.75;red,1}] (.2*\l+\rii,-\l/2-\di,-\l*.85) -- (.2*\l+\rii,-\l/2-\di,-\l*.15);
        \draw[line width=1,color={rgb:black,.75;red,1}] (.2*\l+\rii,-\l*.85-\di,-\l/2) -- (.2*\l+\rii,-\l*.15-\di,-\l/2);
        
        \pgfmathsetmacro{\riii}{\rii+.8*\l}
        \draw[line width=.5] (.45*\l+\riii,-\l*.95-\di,-\l/2) -- ++(0,.9*\l,0);
        \draw[line width=1,color={rgb:black,.75;red,1}] (.45*\l+\riii,-\l/2-\di,-\l*.85) -- (.45*\l+\riii,-\l/2-\di,-\l*.15);
        \draw[line width=1,color={rgb:black,.75;red,1}] (\riii+.125,-\l/2-\di,-\l/2) -- ++(\l*.75,0,0);
        
        \draw (0,0,-\l/2) node[fill=none] {\footnotesize $X$};
        \draw (0,-\l,-\l/2) node[fill=none] {\footnotesize$X$};
        \draw (-\l,0,-\l/2) node[fill=none] {\footnotesize$X$};
        \draw (-\l,-\l,-\l/2) node[fill=none] {\footnotesize$X$};
        \draw (0,-\l*.4,0) node[fill=white] {\footnotesize$X$};
        \draw (0,-\l/2,-\l) node[fill=white] {\footnotesize$X$};
        \draw (-\l,-\l*.6,-\l) node[fill=white] {\footnotesize$X$};
        \draw (-\l,-\l/2,0) node[fill=white] {\footnotesize$X$};
        \draw (-\l*.4,0,0) node[fill=white] {\footnotesize$X$};
        \draw (-\l/2,-\l,0) node[fill=white] {\footnotesize$X$};
        \draw (-\l/2,0,-\l) node[fill=white] {\footnotesize$X$};
        \draw (-\l*.6,-\l,-\l) node[fill=white] {\footnotesize$X$};
        
        \draw (\ri,-\l/2-\di,-\l/2) node[fill=none] {\footnotesize$Z$};
        \draw (\ri+.9*\l,-\l/2-\di,-\l/2) node[fill=none] {\footnotesize$Z$};
        \draw (\ri+.45*\l,-\l/2+.45*\l-\di,-\l/2) node[fill=none] {\footnotesize$Z$};
        \draw (\ri+.45*\l,-\l/2-.45*\l-\di,-\l/2) node[fill=none] {\footnotesize$Z$};
        
        \draw (\riii,-\l/2-\di,-\l/2) node[fill=none] {\footnotesize$Z$};
        \draw (\riii+.9*\l,-\l/2-\di,-\l/2) node[fill=none] {\footnotesize$Z$};
        \draw (\riii+.45*\l,-\l/2-\di,-\l/2+.55*\l) node[fill=none] {\footnotesize$Z$};
        \draw (\riii+.45*\l,-\l/2-\di,-\l/2-.55*\l) node[fill=none] {\footnotesize$Z$};
        
        \draw (\rii+.2*\l,-\l/2+.45*\l-\di,-\l/2) node[fill=none] {\footnotesize$Z$};
        \draw (\rii+.2*\l,-\l/2-.45*\l-\di,-\l/2) node[fill=none] {\footnotesize$Z$};
        \draw (\rii+.2*\l,-\l/2-\di,-\l/2+.55*\l) node[fill=none] {\footnotesize$Z$};
        \draw (\rii+.2*\l,-\l/2-\di,-\l/2-.55*\l) node[fill=none] {\footnotesize$Z$};

    \end{tikzpicture}
    \begin{minipage}{.33\columnwidth}
    {\bf (a)}
    \end{minipage}
    \begin{minipage}{.65\columnwidth}
    {\bf (b)}
    \end{minipage}
    \caption{{\bf (a)} Cube and {\bf (b)} cross operators of the X-cube model Hamiltonian on a cubic lattice.}
    \label{Xc-T3-H}
\end{figure}
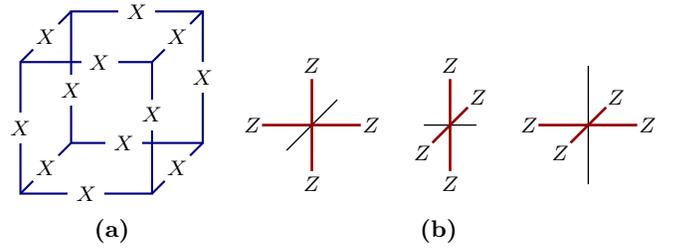

The X-cube model, as first discussed in Ref.~\onlinecite{Sagar16}, is defined on a cubic lattice with qubit degrees of freedom on the edges.
The Hamiltonian
\begin{equation}
H = -\sum_v \left(A_v^{x}+A_v^{y}+A_v^{z}\right) -\sum_c B_c   
\label{eq:H}
\end{equation}
contains two types of terms: cube terms $B_c$ which are products of the twelve Pauli $X$ operators around a cube $c$, and cross terms $A^\mu_v$ which are products of the four Pauli $Z$ operators at a vertex $v$ in the plane normal to the $\mu$-direction where $\mu=x,y,\text{ or }z$ (\figref{Xc-T3-H}).

Consider an $L_x\times L_y\times L_z$ cubic lattice with periodic boundary conditions. The ground state degeneracy (GSD) scales linearly with the size of the system in all three directions:
\begin{equation}
    \log_2{\textrm{GSD}}=2L_x+2L_y+2L_z-3.
    \label{eq:GSDtorus}
\end{equation}

\begin{figure}
    \centering
    \begin{tikzpicture}
        \pgfmathsetmacro{\l}{3}
        \draw[line width=.6]
            (-\l,-\l,0) --++ (0,0,-\l)
            (-\l,0,0) --++ (0,0,-\l)
            (0,-\l,0) --++ (0,0,-\l)
            (0,0,0) --++ (0,0,-\l)
            (0,0,0) -- ++(-\l,0,0) -- ++(0,-\l,0) -- ++(\l,0,0) -- cycle
            (0,0,-\l) -- ++(-\l,0,0) -- ++(0,-\l,0) -- ++(\l,0,0) -- cycle;
        \draw[line width=.4,color=gray]
            (-1*\l/12,-\l,-\l)--++(0,\l,0)
            (-2*\l/12,-\l,-\l)--++(0,\l,0)
            (-3*\l/12,-\l,-\l)--++(0,\l,0)
            (-4*\l/12,-\l,-\l)--++(0,\l,0)
            (-5*\l/12,-\l,-\l)--++(0,\l,0)
            (-6*\l/12,-\l,-\l)--++(0,\l,0)
            (-7*\l/12,-\l,-\l)--++(0,\l,0)
            (-8*\l/12,-\l,-\l)--++(0,\l,0)
            (-9*\l/12,-\l,-\l)--++(0,\l,0)
            (-10*\l/12,-\l,-\l)--++(0,\l,0)
            (-11*\l/12,-\l,-\l)--++(0,\l,0)
        ;
        \draw[line width=.4,color=gray]
            (-\l,-1*\l/12,-\l)--++(\l,0,0)
            (-\l,-2*\l/12,-\l)--++(\l,0,0)
            (-\l,-3*\l/12,-\l)--++(\l,0,0)
            (-\l,-4*\l/12,-\l)--++(\l,0,0)
            (-\l,-5*\l/12,-\l)--++(\l,0,0)
            (-\l,-6*\l/12,-\l)--++(\l,0,0)
            (-\l,-7*\l/12,-\l)--++(\l,0,0)
            (-\l,-8*\l/12,-\l)--++(\l,0,0)
            (-\l,-9*\l/12,-\l)--++(\l,0,0)
            (-\l,-10*\l/12,-\l)--++(\l,0,0)
            (-\l,-11*\l/12,-\l)--++(\l,0,0)
        ;
        \draw[line width=1,color={rgb:blue,1;black,1}]
            (-7*\l/12,-6*\l/12,0)--++(0,0,-\l)
            (-2*\l/12,-6*\l/12,0)--++(0,0,-\l)
            (-7*\l/12,-10*\l/12,0)--++(0,0,-\l)
            (-2*\l/12,-10*\l/12,0)--++(0,0,-\l)
        ;
        \draw[line width=.4,color=gray]
            (-7*\l/12,0,0)--++(0,-\l,0)
            (-2*\l/12,0,0)--++(0,-\l,0)
            (0,-6*\l/12,0)--++(-\l,0,0)
            (0,-10*\l/12,0)--++(-\l,0,0)
        ;
        \draw[line width=.4,color=gray]
            (-1*\l/12,-\l,0)--++(0,\l,0)
            (-2*\l/12,-\l,0)--++(0,\l,0)
            (-3*\l/12,-\l,0)--++(0,\l,0)
            (-4*\l/12,-\l,0)--++(0,\l,0)
            (-5*\l/12,-\l,0)--++(0,\l,0)
            (-6*\l/12,-\l,0)--++(0,\l,0)
            (-7*\l/12,-\l,0)--++(0,\l,0)
            (-8*\l/12,-\l,0)--++(0,\l,0)
            (-9*\l/12,-\l,0)--++(0,\l,0)
            (-10*\l/12,-\l,0)--++(0,\l,0)
            (-11*\l/12,-\l,0)--++(0,\l,0)
        ;
        \draw[line width=.4,color=gray]
            (-\l,-1*\l/12,0)--++(\l,0,0)
            (-\l,-2*\l/12,0)--++(\l,0,0)
            (-\l,-3*\l/12,0)--++(\l,0,0)
            (-\l,-4*\l/12,0)--++(\l,0,0)
            (-\l,-5*\l/12,0)--++(\l,0,0)
            (-\l,-6*\l/12,0)--++(\l,0,0)
            (-\l,-7*\l/12,0)--++(\l,0,0)
            (-\l,-8*\l/12,0)--++(\l,0,0)
            (-\l,-9*\l/12,0)--++(\l,0,0)
            (-\l,-10*\l/12,0)--++(\l,0,0)
            (-\l,-11*\l/12,0)--++(\l,0,0)
        ;
        \draw[line width=1,color={rgb:green,.5;black,1}]
            (-10*\l/12,-3*\l/12,0)--++(0,0,-\l)
        ;
        
        \draw[arrows={-latex}] (\l*.4,-\l,-\l) -- ++(\l*.4,0,0);
        \draw[arrows={-latex}] (\l*.4,-\l,-\l) -- ++(0,\l*.4,0);
        \draw[arrows={-latex}] (\l*.4,-\l,-\l) -- ++(0,0,\l/1.7);
        
        \draw (\l*.4+.2,-\l+1.2,-\l) node[fill=none] {$y$};
        \draw (\l-.6,-\l-.2,-\l) node[fill=none] {$x$};
        \draw (\l*.4-.7,-\l-.4,-\l) node[fill=none] {$z$};
        
        \draw(-\l/12*7,-\l*1.07,0) node[fill=none] {$i$};
        \draw(-\l/12*2,-\l*1.07,0) node[fill=none] {$j$};
        \draw(-\l/12*10,-\l*1.07,0) node[fill=none] {$m$};
        
        \draw(-\l*1.06,-\l/12*3,0) node[fill=none] {$n$};
        \draw(-\l*1.06,-\l/12*6,0) node[fill=none] {$k$};
        \draw(-\l*1.06,-\l/12*10,0) node[fill=none] {$l$};
    \end{tikzpicture}
    \caption{Visualization of logical operators in the X-cube model. The green string corresponds to $W^z_{mn}$. The product of the four operators corresponding to the blue strings is equal to the identity, as described in the main text.}
    \label{fig:logicalOperators}
\end{figure}
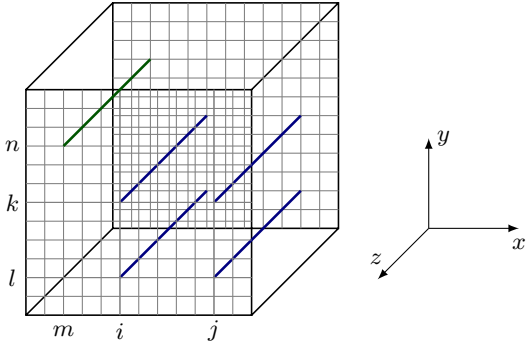

There are hence a large number of `logical operators' that commute with all of the terms in the Hamiltonian and map one ground state to another.\cite{Slagle17QFT,BernevigEntropy}
An over-complete set of $X$-type logical operators is given by the set of closed string-like operators $W^\mu_{ij}$,
which is a product of $X$ operators over all $\mu$-oriented edges with coordinates $(i,j)$ in the plane normal to $\mu$ (see \figref{fig:logicalOperators}).
This set is over-complete in the sense that products of the form $W^\mu_{ij}W^\mu_{il}W^\mu_{kl}W^\mu_{kj}$ are equal to a product of some $B_c$ cube operators, and thus act trivially on the ground state manifold (here the four sets of coordinates lie ahbt the corners of a rectangle in the plane normal to $\mu$, as shown in \figref{fig:logicalOperators}).
There are $L_xL_y+L_yL_z+L_zL_x-2L_x-2L_y-2L_z+3$ such relations corresponding to unique products of cube operators, thus implying \eqnref{eq:GSDtorus}. 

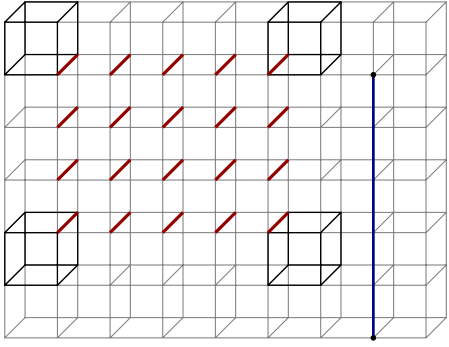
\begin{figure}
    \centering
    \begin{tikzpicture}
        \pgfmathsetmacro{\l}{.7}
        \draw[line width=.4, color=gray]
            (0,0,0) --++ (0,0,-\l)
            (0,-1*\l,0) --++ (0,0,-\l)
            (0,-2*\l,0) --++ (0,0,-\l)
            (0,-3*\l,0) --++ (0,0,-\l)
            (0,-4*\l,0) --++ (0,0,-\l)
            (0,-5*\l,0) --++ (0,0,-\l)
            (-1*\l,0,0) --++ (0,0,-\l)
            (-1*\l,-1*\l,0) --++ (0,0,-\l)
            (-1*\l,-2*\l,0) --++ (0,0,-\l)
            (-1*\l,-3*\l,0) --++ (0,0,-\l)
            (-1*\l,-4*\l,0) --++ (0,0,-\l)
            (-1*\l,-5*\l,0) --++ (0,0,-\l)
            (-2*\l,0,0) --++ (0,0,-\l)
            (-2*\l,-1*\l,0) --++ (0,0,-\l)
            (-2*\l,-2*\l,0) --++ (0,0,-\l)
            (-2*\l,-3*\l,0) --++ (0,0,-\l)
            (-2*\l,-4*\l,0) --++ (0,0,-\l)
            (-2*\l,-5*\l,0) --++ (0,0,-\l)
            (-3*\l,0,0) --++ (0,0,-\l)
            (-3*\l,-1*\l,0) --++ (0,0,-\l)
            (-3*\l,-2*\l,0) --++ (0,0,-\l)
            (-3*\l,-3*\l,0) --++ (0,0,-\l)
            (-3*\l,-4*\l,0) --++ (0,0,-\l)
            (-3*\l,-5*\l,0) --++ (0,0,-\l)
            (-4*\l,0,0) --++ (0,0,-\l)
            (-4*\l,-1*\l,0) --++ (0,0,-\l)
            (-4*\l,-2*\l,0) --++ (0,0,-\l)
            (-4*\l,-3*\l,0) --++ (0,0,-\l)
            (-4*\l,-4*\l,0) --++ (0,0,-\l)
            (-4*\l,-5*\l,0) --++ (0,0,-\l)
            (-5*\l,0,0) --++ (0,0,-\l)
            (-5*\l,-1*\l,0) --++ (0,0,-\l)
            (-5*\l,-2*\l,0) --++ (0,0,-\l)
            (-5*\l,-3*\l,0) --++ (0,0,-\l)
            (-5*\l,-4*\l,0) --++ (0,0,-\l)
            (-5*\l,-5*\l,0) --++ (0,0,-\l)
            (-6*\l,0,0) --++ (0,0,-\l)
            (-6*\l,-1*\l,0) --++ (0,0,-\l)
            (-6*\l,-2*\l,0) --++ (0,0,-\l)
            (-6*\l,-3*\l,0) --++ (0,0,-\l)
            (-6*\l,-4*\l,0) --++ (0,0,-\l)
            (-6*\l,-5*\l,0) --++ (0,0,-\l)
            (-7*\l,0,0) --++ (0,0,-\l)
            (-7*\l,-1*\l,0) --++ (0,0,-\l)
            (-7*\l,-2*\l,0) --++ (0,0,-\l)
            (-7*\l,-3*\l,0) --++ (0,0,-\l)
            (-7*\l,-4*\l,0) --++ (0,0,-\l)
            (-7*\l,-5*\l,0) --++ (0,0,-\l)
            
            (0,0,0)--++(-\l*8,0,0)
            (0,-1*\l,0)--++(-\l*8,0,0)
            (0,-2*\l,0)--++(-\l*8,0,0)
            (0,-3*\l,0)--++(-\l*8,0,0)
            (0,-4*\l,0)--++(-\l*8,0,0)
            (-1*\l,\l,0)--++(0,-\l*6,0)
            (-2*\l,\l,0)--++(0,-\l*6,0)
            (-3*\l,\l,0)--++(0,-\l*6,0)
            (-4*\l,\l,0)--++(0,-\l*6,0)
            (-5*\l,\l,0)--++(0,-\l*6,0)
            (-6*\l,\l,0)--++(0,-\l*6,0)
            (-7*\l,\l,0)--++(0,-\l*6,0)
            (0,0,-\l)--++(-\l*7,0,0)
            (0,-1*\l,-\l)--++(-\l*8,0,0)
            (0,-2*\l,-\l)--++(-\l*8,0,0)
            (0,-3*\l,-\l)--++(-\l*8,0,0)
            (0,-4*\l,-\l)--++(-\l*8,0,0)
            (-1*\l,\l,-\l)--++(0,-\l*6,0)
            (-2*\l,\l,-\l)--++(0,-\l*6,0)
            (-3*\l,\l,-\l)--++(0,-\l*6,0)
            (-4*\l,\l,-\l)--++(0,-\l*6,0)
            (-5*\l,\l,-\l)--++(0,-\l*6,0)
            (-6*\l,\l,-\l)--++(0,-\l*6,0)
            (-7*\l,\l,-\l)--++(0,-\l*6,0)
            
            (-0*\l,\l,0)--++(0,0,-\l)
            (-1*\l,\l,0)--++(0,0,-\l)
            (-2*\l,\l,0)--++(0,0,-\l)
            (-3*\l,\l,0)--++(0,0,-\l)
            (-4*\l,\l,0)--++(0,0,-\l)
            (-5*\l,\l,0)--++(0,0,-\l)
            (-6*\l,\l,0)--++(0,0,-\l)
            (-7*\l,\l,0)--++(0,0,-\l)
            (-8*\l,\l,0)--++(0,0,-\l)
            (-8*\l,0*\l,0)--++(0,0,-\l)
            (-8*\l,-1*\l,0)--++(0,0,-\l)
            (-8*\l,-2*\l,0)--++(0,0,-\l)
            (-8*\l,-3*\l,0)--++(0,0,-\l)
            (-8*\l,-4*\l,0)--++(0,0,-\l)
            (-8*\l,-5*\l,0)--++(0,0,-\l)
            
            (0,\l,0)--++(-8*\l,0,0)--++(0,-6*\l,0)--++(8*\l,0,0)--cycle
            (0,\l,-\l)--++(-8*\l,0,0)--++(0,-6*\l,0)--++(8*\l,0,0)--cycle
            ;
            
            \draw[line width=.5]
            (-2*\l,+\l,0)--++(-\l,0,0)--++(0,-\l,0)--++(\l,0,0)--cycle
            (-7*\l,+\l,0)--++(-\l,0,0)--++(0,-\l,0)--++(\l,0,0)--cycle
            (-2*\l,+\l,-\l)--++(-\l,0,0)--++(0,-\l,0)--++(\l,0,0)--cycle
            (-7*\l,+\l,-\l)--++(-\l,0,0)--++(0,-\l,0)--++(\l,0,0)--cycle
            (-2*\l,-3*\l,0)--++(-\l,0,0)--++(0,-\l,0)--++(\l,0,0)--cycle
            (-7*\l,-3*\l,0)--++(-\l,0,0)--++(0,-\l,0)--++(\l,0,0)--cycle
            (-2*\l,-3*\l,-\l)--++(-\l,0,0)--++(0,-\l,0)--++(\l,0,0)--cycle
            (-7*\l,-3*\l,-\l)--++(-\l,0,0)--++(0,-\l,0)--++(\l,0,0)--cycle
            (-8*\l,\l,0)--++(0,0,-\l)
            (-8*\l,0,0)--++(0,0,-\l)
            (-7*\l,\l,0)--++(0,0,-\l)
            (-2*\l,\l,0)--++(0,0,-\l)
            (-2*\l,0,0)--++(0,0,-\l)
            (-3*\l,\l,0)--++(0,0,-\l)
            (-8*\l,-3*\l,0)--++(0,0,-\l)
            (-8*\l,-4*\l,0)--++(0,0,-\l)
            (-7*\l,-4*\l,0)--++(0,0,-\l)
            (-2*\l,-3*\l,0)--++(0,0,-\l)
            (-2*\l,-4*\l,0)--++(0,0,-\l)
            (-3*\l,-4*\l,0)--++(0,0,-\l)
            ;
            
            \draw[line width=1.2,color={rgb:red,1;black,.75}]
            (-\l*6,-\l*1,0)--++(0,0,-\l)
            (-\l*5,-\l*1,0)--++(0,0,-\l)
            (-\l*4,-\l*1,0)--++(0,0,-\l)
            (-\l*3,-\l*1,0)--++(0,0,-\l)
            (-\l*7,-\l*1,0)--++(0,0,-\l)
            (-\l*6,-\l*2,0)--++(0,0,-\l)
            (-\l*5,-\l*2,0)--++(0,0,-\l)
            (-\l*4,-\l*2,0)--++(0,0,-\l)
            (-\l*3,-\l*2,0)--++(0,0,-\l)
            (-\l*7,-\l*2,0)--++(0,0,-\l)
            (-\l*6,-\l*3,0)--++(0,0,-\l)
            (-\l*5,-\l*3,0)--++(0,0,-\l)
            (-\l*4,-\l*3,0)--++(0,0,-\l)
            (-\l*3,-\l*3,0)--++(0,0,-\l)
            (-\l*7,-\l*3,0)--++(0,0,-\l)
            (-\l*6,0,0)--++(0,0,-\l)
            (-\l*5,0,0)--++(0,0,-\l)
            (-\l*4,0,0)--++(0,0,-\l)
            (-\l*3,0,0)--++(0,0,-\l)
            (-\l*7,0,0)--++(0,0,-\l)
            ;
            
            \draw[line width=1,color={rgb:blue,1;black,1}]
            (-\l*1,-\l*5,0)--++(0,\l*5,0)
            ;
            
            \draw[black,fill=black] (-\l,0,0) circle (.03);
            \draw[black,fill=black] (-\l,-\l*5,0) circle (.03);
            
    \end{tikzpicture}
    \caption{Visualization of particle creation operators in the X-cube model. The red links correspond to a membrane geometry on the dual lattice. The product of $Z$ operators over these edges excites the (darkened) cube operators at the corners. The product of $X$ operators over the links comprising the straight open blue string creates excitations at its endpoints (black dots).}
    \label{fig:excitations}
\end{figure}

Logical operators correspond to processes where particle anti-particle pairs are created out of the vacuum, wound around the torus, and then annihilated. Straight open string operators $W^\mu_{ij}\left(\mu_1,\mu_2\right)$ anti-commute with the vertex Hamiltonian terms at the endpoints $\mu_1$ and $\mu_2$, corresponding to excitations which live on the vertices of the lattice. Here $W^\mu_{ij}\left(\mu_1,\mu_2\right)$ is defined to be the product of $X$ operators over $\mu$-oriented edges between $\mu=\mu_1$ and $\mu=\mu_2$ with coordinate $(i,j)$ in the plane normal to $\mu$ (see \figref{fig:excitations}). Conversely, acting with bent string operators introduces additional energetic costs at the corners. Therefore the particles living at the endpoints of straight open strings are energetically confined to live on a line; in this sense, they are dimension-1 particles. \cite{Sagar16} These particles obey an unconventional fusion rule: triples of particles living along $x$-, $y$-, and $z$-oriented lines may annihilate into the vacuum. On the other hand, acting with a closed string 
operator around a rectangle creates an excitation at each corner of the rectangle. A pair of particles at adjacent corners may be viewed as a single dipole-like object which is itself a dimension-2 particle and is mobile in the plane normal to the edges connecting the two corners.

In addition to these string-like operators, there are membrane-like operators which are products of $Z$ operators over qubits corresponding to a membrane geometry on the dual lattice (see \figref{fig:excitations}). A rectangular membrane operator anti-commutes with the cube Hamiltonian terms at its corners. A pair of adjacent corner excitations created by a rectangular membrane operator is likewise a dimension-2 dipolar particle, free to move in a plane perpendicular to its moment. A process whereby a pair of such membrane dipoles is created, separated, wound around the torus and annihilated, corresponds to a string-like $Z$-type logical operator.

\section{Ground state degeneracy}
\label{app:GSD}

In this appendix, we review algorithms to compute the ground state degeneracy and entanglement entropy of a $Z_D$ qudit stabilizer code. \cite{GheorghiuQudit,FattalEntanglement,Gottesman97}

Consider a stabilizer code of the form
\begin{align}
  H &= - \sum_{\alpha=1}^k \, (s_\alpha + s_\alpha^\dagger), \\
  s_\alpha &= \omega^{p_\alpha} \prod_{i=1}^n X_i^{S_{\alpha,i}} Z_i^{S_{\alpha,i+n}}.
\end{align}
Each $s_\alpha$ is a product of $Z_D$ clock and shift operators $Z$ and $X$ where $ZX=\omega XZ$ and $\omega = e^{2\pi i/D}$.
Note that $H$ is completely determined by the $k$-component integer vector $p_\alpha$ and $k \times 2n$ integer matrix $S$.
Since we require that $H$ is a stabilizer code,
  any product of $s_\alpha$ that results in a multiple of the identity operator must be the identity operator exactly;
  i.e. $H$ must be frustration-free.

Multiplying one stabilizer by another or applying unitary Clifford operators to $H$ roughly corresponds to multiplying $S$ on the left or right by an invertible integer matrix,
  along with some additional modifications to $p_\alpha$.
Analogous to the singular value decomposition, the Smith decomposition diagonalizes an integer matrix using invertible integer matrices.
Therefore, we can compute the Smith normal form of $S$ to obtain a new integer matrix $S'$ which is diagonal,
  and the Hamiltonian $H'$ defined by $S'$ will have the same ground state degeneracy as $H$.
Since $S'$ is diagonal, $H'$ consists of decoupled qudits,
  and the ground state degeneracy of $H'$ is trivial to calculate
  (and the new phases $\omega^{p'_\alpha}$ do not affect the degeneracy).
For the special case of $Z_D$ qudits with $D$ prime,
  the degeneracy can instead be calculated from the rank of $S$ over the field $Z_D$.

An algorithm to compute the entanglement of a qubit stabilizer code is discussed in \refcite{FattalEntanglement}.
Similar to the ground state degeneracy calculation,
  the entanglement entropy is computed in terms of the rank of a matrix $S_{AB}$ over the field $Z_D$ when the qudit dimension $D$ is prime.
For non-prime $D$, the algorithm generalizes similarly to the degeneracy calculation and the entanglement entropy is calculated from the Smith diagonals of the same matrix.

\section{String-membrane-net realization}
\label{sec:SMN}

\begin{figure}
    \centering
    \includegraphics[width=.45\textwidth]{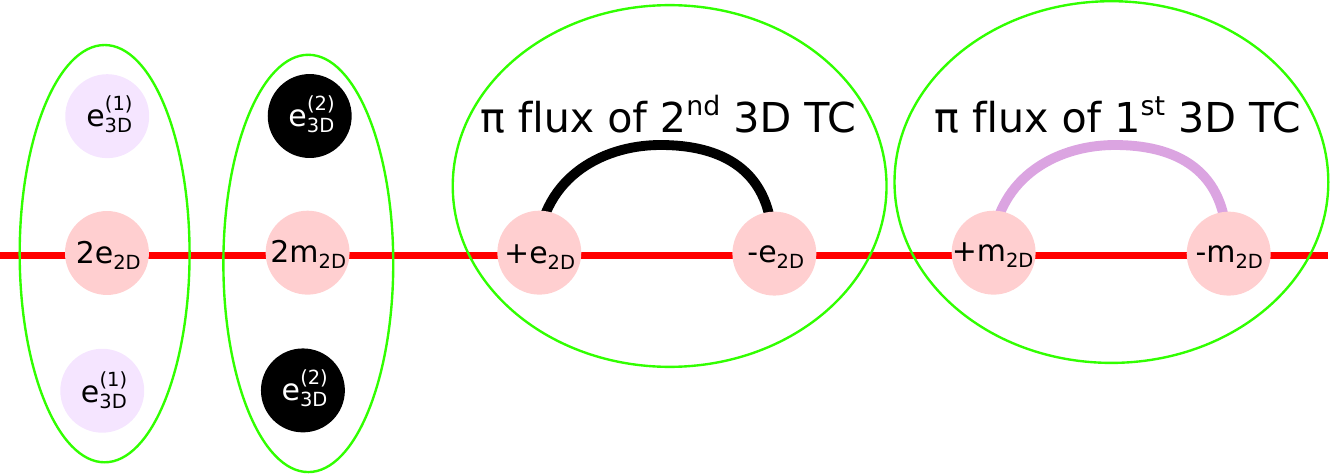}
    \caption{
    The four kinds of excitations (circled in green) that can be created locally in the 3-foliated string-membrane-net.
    }
    \label{fig:condensed}
\end{figure}

The 3-foliated model in \secref{sec:3foliatedCSS} can also be written as a string-membrane-net (SMN) model \cite{SlagleSMN}.
The SMN consists of \emph{two} 3D $Z_2$ toric codes coupled to 2D $Z_4$ toric code (TC) layers.
The coupling modifies the set of local excitations along the 2D layers, which in turn modifies the mobility of the excitations:
\begin{enumerate}
  \item When a pair of charges $e_\text{3D}^\text{(1)}$ ($e_\text{3D}^\text{(2)}$) of the first (or second) 3D TC is created across a layer,
    a pair of charge $2e_\text{2D}$ (flux $2m_\text{2D}$) excitations is also created on the 2D TC layer.
  \item When a pair of oppositely-charged $Z_4$ charge $\pm e_\text{2D}$ (or flux $\pm m_\text{2D}$) excitations is created on a 2D TC layer,
    an open $\pi$ flux string excitation of the second (first) 3D TC is also created with endpoints on the two oppositely-charged 2D excitations.
\end{enumerate}
See \figref{fig:condensed} for pictures of these local excitations.

Note that the mobility of particles is determined by the set of local excitations since charges can move by creating and annihilating local excitations,
  such as a pair of slightly displaced excitations of opposite charge.
However, exotic sets of local excitations lead to more interesting mobility rules.
For example, due to the first effect above,
  the 3D toric code (TC) charges ($e_\text{3D}^{(1)}$ and $e_\text{3D}^{(2)}$) are fractons since they must leave behind 2D TC excitations when they pass through layers.
The second effect implies that an odd number of 2D TC charges ($e_\text{(2D)}$) or fluxes ($m_\text{(2D)}$) must be attached to the endpoints of 3D TC flux strings,
  which implies that an odd number of 2D TC charges or fluxes are linearly confined.
However, a pair of 2D TC charges (or fluxes) from two intersecting layers is a lineon because this pair is confined to the intersection of the two layers by the 3D TC flux strings.

\begin{figure*}
    \centering
    \begin{tabular}{ccccc}
      2D flux $m_\text{2D}$ & 2D charge $e_\text{2D}$ & 1\textsuperscript{st} 3D flux & 2\textsuperscript{nd} 3D flux & 1\textsuperscript{st} 3D charge $e_\text{3D}^{(1)}$ \\
      $A_e^\text{(SMN)}=$ & $B_p^\text{(SMN)}=$ & $C_e^\text{(SMN)}=$ & $D_p^\text{(SMN)}=$ & $E_v^\text{(SMN)}=$ \\
        $\tilde{Z}\tilde{Z}\tilde{Z}^\dagger\tilde{Z}^\dagger \tau^z$
      & $\tilde{X}\tilde{X}\tilde{X}^\dagger\tilde{X}^\dagger \sigma^x$
      & $\sigma^z\sigma^z\sigma^z\sigma^z\tilde{Z}^2\tilde{Z}^2$
      & $\tau^x\tau^x\tau^x\tau^x\tilde{X}^2\tilde{X}^2$
      & $\sigma^x\sigma^x\sigma^x\sigma^x\sigma^x\sigma^x$ \\
      \opic{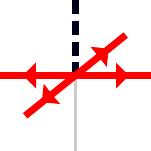} & \opic{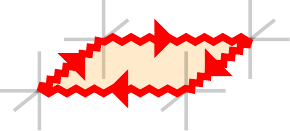} & \opic{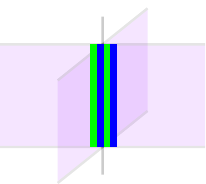} & \opic{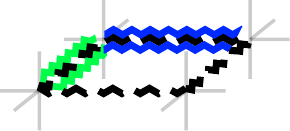} & \opic{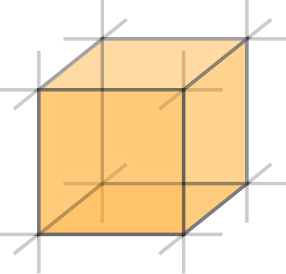} \\
      \opic{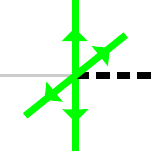} & \opic{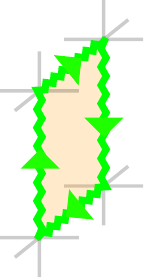} & \opic{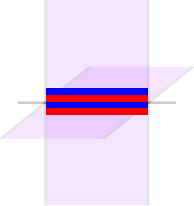} & \opic{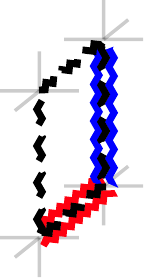}
      & \begin{tabular}{c} \\ \\ \\ 2\textsuperscript{nd} 3D charge $e_\text{3D}^{(2)}$ \\ $F_c^\text{(SMN)}=$ \\
        $\tau^z\tau^z\tau^z\tau^z\tau^z\tau^z$ \end{tabular} \\
      \opic{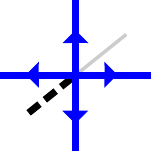} & \opic{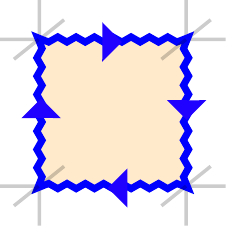} & \opic{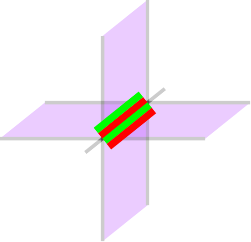} & \opic{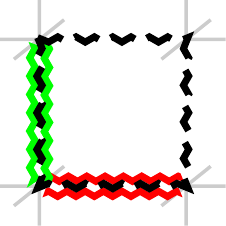} & \opic{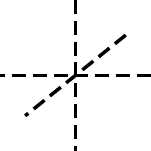}
    \end{tabular}
    \caption{
    A depiction of the terms in the string-membrane-net Hamiltonian $H = - \sum_e A_e^\text{(SMN)} - \sum_p B_p^\text{(SMN)} - \sum_e C_e^\text{(SMN)} - \sum_p D_p^\text{(SMN)} - \sum_v E_v^\text{(SMN)} - \sum_c F_c^\text{(SMN)}$.
    The Hamiltonian consists of three stacks of $Z_4$ 2D toric codes coupled to two $Z_2$ 3D toric codes.
    The 2D toric codes consist of $Z_4$ qudits on the edges of stacks of 2D square lattices.
    The operators of the 2D toric codes on the $xy$, $yz$, and $zx$ planes will be colored red, green, and blue.
    A straight red, green, or blue line denotes a $Z_4$ clock operator $\tilde{Z}$,
      while a zig-zag line denotes a $Z_4$ shift operator $\tilde{X}$ with the algebra $\tilde{Z}\tilde{X}=i\tilde{X}\tilde{Z}$.
    When e.g. two red lines appear on the same edge, this denotes a $\tilde{Z}^2$ operator.
    A conjugate-transpose is taken for operators on edges with arrows that point in the negative $x$, $y$, or $z$ direction.
    The first 3D toric code consists of $Z_2$ qubits on the plaquettes of the cubic lattice,
      for which purple and orange plaquettes denote $Z_2$ Pauli $\sigma^z$ and $\sigma^x$ operators, respectively.
    The second 3D toric code consists of $Z_2$ qubits on the links of the cubic lattice,
      which are denoted by dashed back lines;
      again, straight and zig-zag lines denote Pauli $\tau^z$ and $\tau^x$ operators.
    Thus, there are two $Z_4$ qudits and one $Z_2$ qubit on each edge,
      and a single $Z_2$ qubit on each plaquette.
    The Hamiltonian consists of these 14 different operators, along with their Hermitian conjugates.
    Above each column of operators, are written the name of the corresponding excitation and the individual Pauli, clock, and shift operators that the operators are composed of.}
    \label{fig:SMN}
\end{figure*}

The Hamiltonian of the string-membrane-net can be written down on very general lattices.
In particular, it is possible to consider lattices where there are many qubits between the toric code layers so that one can indeed think of the Hamiltonian as 2D toric codes coupled to two 3D toric codes.
In \figref{fig:SMN}, we depict the simplest example where the Hamiltonian is defined on a cubic lattice in which the toric code layers are placed a single lattice spacing apart from one another.

\subsection{Unitary Mapping}

To show that the string-membrane-net Hamiltonian (\figref{fig:SMN}) is equivalent to the cage-net Hamiltonian in \figref{fig:CSS cagenet},
  we will show that there is a unitary mapping between the ground spaces of the two Hamiltonians (augmented with some extra decoupled degrees of freedom).

To begin, it is convenient to replace the $Z_2$ qubits of the two 3D toric codes with $Z_4$ qudits.
This will be achieved by making the following operator replacement in the string-membrane-net Hamiltonian (\figref{fig:SMN}):
\begin{align}
  \sigma_p^z &\to (\tilde\sigma_p^z)^2 & \sigma_p^x &\to \tilde\sigma_p^x \label{eq:Z2 to Z4}\\
  \tau_e^x   &\to (\tilde\tau_e^x  )^2 & \tau_e^z   &\to \tilde\tau_e^z \nonumber
\end{align}
and adding the following terms to the Hamiltonian:
\begin{align}
  -\sum_p (\tilde\sigma_p^x)^2 - \sum_e (\tilde\tau_e^z)^2 \label{eq:H Z2 to Z4}
\end{align}
We have replaced the Pauli operators $\sigma^\mu$ and $\tau^\mu$ with clock and shift operators $\tilde\sigma^\mu$ and $\tilde\tau^\mu$,
  which have the algebra $\tilde\sigma^z \tilde\sigma^x = i \tilde\sigma^x \tilde\sigma^z$ and $\tilde\tau^z \tilde\tau^x = i \tilde\tau^x \tilde\tau^z$.
The above replacement does not change the ground state since the new terms in the Hamiltonian will enforce $\tilde\sigma=\pm1$ and $\tilde\tau=\pm1$,
  and the modified Hamiltonian does not have any $\sigma^z$ or $\tau^x$ operators, only $(\tilde\sigma^z)^2$ and $(\tilde\tau^x)^2$ operators.
Thus, its ground state is still effectively described by qubits.

\begin{figure}
    \centering
    \begin{tabular}{cc}
      \opic{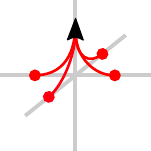} & \opic{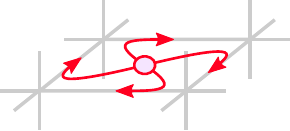} \\
      \opic{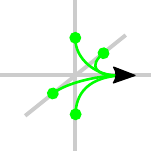} & \opic{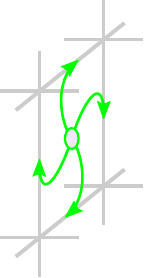} \\
      \opic{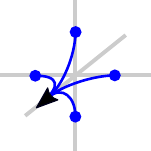} & \opic{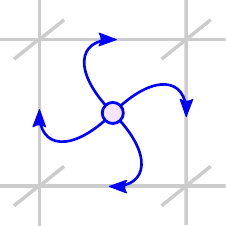}
    \end{tabular}
    \caption{After applying the mapping in \eqnref{eq:Z2 to Z4},
      the unitary depicted above maps the string-membrane-net model in \figref{fig:SMN} to the cage-net Hamiltonian in \figref{fig:CSS cagenet}.
      The unitary is given by the composition of a unitary operator at each edge (left) and plaquette (right).
      These smaller unitary operators commute with each other.
      The operators on the left are products of four controlled-X operators (one for each line, defined in \eqnref{eq:CX}) that are controlled by the 2D toric code qudit of the appropriate color at the colored dot,
        and act on the 3D toric code qudit at the end of the black arrow.
      The operators on the right are controlled-X operators that are controlled by the 3D toric code qudit at the center of the plaquette,
        and act on the 2D toric code qudit of the appropriate color at the end of the arrow.}
    \label{fig:SMN unitary}
\end{figure}

The next step is to act with the unitary shown in \figref{fig:SMN unitary},
  which is composed of the $Z_4$ controlled-X operators:
\begin{gather}
  CX = \frac{1}{4} \sum_{a=0}^3 \sum_{b=0}^3 i^{ab} Z^a \otimes X^b \label{eq:CX},\\
\begin{aligned}
  CX (Z \otimes 1) CX^\dagger &= Z \otimes 1, & CX (X \otimes 1) CX^\dagger &= X \otimes X^{-1} \\
  CX (1 \otimes Z) CX^\dagger &= Z \otimes Z, & CX (1 \otimes X) CX^\dagger &= 1 \otimes X.
\end{aligned} \nonumber
\end{gather}

The replacement in \eqnref{eq:Z2 to Z4} and unitary in \figref{fig:SMN unitary}
  map the operators of the string-membrane-net Hamiltonian (\figref{fig:SMN}) to those of the cage-net Hamiltonian (\figref{fig:CSS cagenet}) as follows
\begin{align}
    A_e^\text{(SMN)} &\to \tau_e^z,
  & C_e^\text{(SMN)} &\to C_e^\text{(cage)},
  & E_v^\text{(SMN)} &\to E_p^\text{(cage)} \nonumber\\
    B_p^\text{(SMN)} &\to \sigma_p^x,
  & D_p^\text{(SMN)} &\to D_p^\text{(cage)},
  & F_v^\text{(SMN)} &\to F_v^\text{(cage)}.
\end{align}
The $A_e^\text{(SMN)}$ and $B_p^\text{(SMN)}$ operators are mapped to $\tau_e^z$ and $\sigma_p^x$.
This sets $\tau_e^z = \sigma_p^x = 1$ in the ground state of the new Hamiltonian.
We also had to add two new terms to the Hamiltonian in \eqnref{eq:H Z2 to Z4}.
These new terms are mapped to
\begin{align}
  (\tilde{\tau}_e^z)^2 &\to (\tilde{\tau}_e^z)^2 A_e^\text{(cage)} &
  (\tilde{\sigma}_p^x)^2 &\to (\tilde{\sigma}_p^x)^2 B_p^\text{(cage)}
\end{align}
But since $\tau_e^z = \sigma_p^x = 1$ in the ground state,
  the new terms are effectively mapped to $A_e^\text{(cage)}$ and $B_p^\text{(cage)}$.
Therefore, the string-membrane-net Hamiltonian (\figref{fig:SMN})
  and cage-net Hamiltonian (\figref{fig:CSS cagenet}) both have the same ground state (up to trivial decoupled degrees of freedom).

\subsection{Field theory}

It is also possible to describe this model using a foliated field theory.
Foliated field theories, which were introduced in \refcite{SlagleSMN},
  are field theories that explicitly couple to a foliation structure via foliation fields $e_\mu^k$.\footnote{The X-cube field theory in \refcite{Slagle17QFT} was written as a foliated field theory in \refcite{SlagleSMN}.}

The Lagrangian is
\begin{align}
  L &= \overbrace{\frac{4}{2\pi} \sum_k e^k \wedge B^k \wedge dA^k}^\text{2D $Z_4$ TC layers}
     + \overbrace{ \frac{2}{2\pi} b \wedge da
     + \frac{2}{2\pi} b' \wedge da'}^\text{2$\times$ 3D $Z_2$ TC} \nonumber\\
    &- \underbrace{\frac{4}{2\pi} \sum_k e^k \wedge (b \wedge A^k + a' \wedge B^k)}_\text{coupling} \label{eq:SMN L}
\end{align}
where $A^k$, $B^k$, $a$, and $b'$ are 1-form gauge fields,
  $b$ and $a'$ are 2-form gauge fields,
  $e^k$ are static foliation fields that describe the geometry of the foliations,
  and $k=1,2,..,n_f$ indexes the different foliation layers.
The $n_f=3$ foliation structure of a cubic lattice is described by
  $e^k_\mu = \lambda \delta^k_\mu$ where $\mu=0,1,2,3$ indexes the spacetime indices and $\lambda$ is the density of foliation layers.

The Lagrangian has the following gauge invariance
\begin{align}
  A^k &\to A^k + d \zeta^k + \alpha' &
  B^k &\to B^k + d \chi^k + \beta \nonumber\\
     &\; + \mu^k e^k &
     &\; + \nu^k e^k \\
  a &\to a + d \alpha - \sum_k 2\zeta^k e^k &
  b &\to b + d \beta \nonumber\\
  a' &\to a' + d \alpha' &
  b' &\to b' + d \beta' - \sum_k 2\chi^k e^k \nonumber
\end{align}
where $\zeta^k$, $\chi^k$, $\mu^k$ $\nu^k$, $\alpha$, and $\beta'$ are arbitrary scalars and
  $\beta$ and $\alpha'$ are arbitrary 1-forms.
The Lagrangian is also self-dual under
\begin{align}
  A^k &\leftrightarrow B^k &
  a   &\leftrightarrow b' &
  a'  &\leftrightarrow b.
\end{align}
This self-duality interchanges the two 3D toric codes and interchanges the 2D toric code charge and flux sectors.
  
\subsection{1-foliated model}
\label{app:1foliatedLattice}

\begin{figure}
    \centering
    \begin{tabular}{cccc}
      2D flux & 2D charge & \multicolumn{2}{c}{3D flux} \\
        $\tilde{Z}\tilde{Z}\tilde{Z}^\dagger\tilde{Z}^\dagger$
      & $\tilde{X}\tilde{X}\tilde{X}^\dagger\tilde{X}^\dagger \sigma^x$
      & $\sigma^z\sigma^z\sigma^z\sigma^z$
      & $\sigma^z\sigma^z\sigma^z\sigma^z\tilde{Z}^2$ \\
      \opic{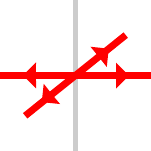} & \opic{Bpxy.pdf} & \opic{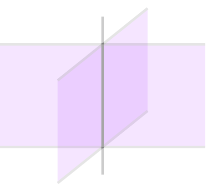} & \opic{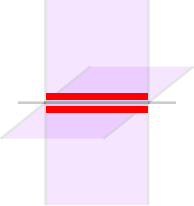} \\
      & \multicolumn{2}{c}{3D charge} & \multirow{3}{*}{\opic{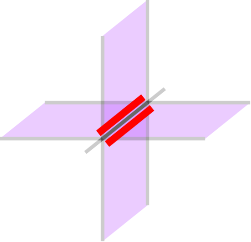}} \\
      & \multicolumn{2}{c}{$\sigma^x\sigma^x\sigma^x\sigma^x\sigma^x\sigma^x$} & \\
      & \multicolumn{2}{c}{\opic{Bcube.pdf}} &
    \end{tabular}
    \caption{A depiction of the terms in the string-membrane-net Hamiltonian realization of the 1-foliated K-matrix in \eqnref{eq:K}.
    The Hamiltonian consists of a single stack of $Z_4$ toric codes coupled to a $Z_2$ 3D toric code.
    The 2D toric codes consist of $Z_4$ qudits on the edges of a stack of 2D square lattices.
    The pictoral notation is similar to that of \figref{fig:SMN}.
    A straight red line denotes a $Z_4$ clock operator $\tilde{Z}$, while a zig-zag line denotes a $Z_4$ shift operator $\tilde{X}$ with the algebra $\tilde{Z} \tilde{X} = i \tilde{X} \tilde{Z}$.
    The 3D toric code consists of $Z_2$ qubits on the plaquettes of the cubic lattice,
      for which purple and orange operators denote $Z_2$ Pauli $\sigma^z$ and $\sigma^x$ operators, respectively.
    Thus, there qre two $Z_4$ qudits on each $x$-axis or $y$-axis edge, no qudits on the $z$-axis edges, and a single $Z_2$ qubit on each plaquette.}
    \label{fig:1fSMN}
\end{figure}

In this appendix, we write down a CSS code lattice model that can describe the twisted 1-foliated K-matrix model in \eqnref{eq:K}.
One option would be to consider the 1-foliated version of the string-membrane-net model in \figref{fig:SMN}.
This appears to work, but the second toric code does not have any affect in this 1-foliated case.
Thus, we will consider the simpler case of a stack of 2D $Z_4$ toric codes coupled to a 3D $Z_2$ toric code.
This model is a special case of the generalized string-membrane-net model in Appendix A of \refcite{SlagleSMN}.
The model is summarized in \figref{fig:1fSMN}.

The anyon labels in \eqnref{eq:K} have the following correspondence with the excitations of the 1-foliated string-membrane-net:
\begin{center}
\begin{tabular}{cc}
 K-matrix anyon & string-membrane-net \\\hline
 $e_{2z+1}$ & pair of 2D fluxes \\
 $m_{2z+1}$ & 2D charge \\
 $e_{2z+2}$ & 3D charge \\
 $m_{2z+2}$ & 2D fluxes - 3D flux - 2D flux \\
\end{tabular}
\end{center}
The anyon $m_{2z+2}$ is equivalent to a pair of 2D fluxes on neighboring layers where the fluxes are attached to two ends of a 3D flux string.
It is straightforward to check that the above anyons have the same braiding statistics as those defined in the $K^{-1}$ matrix in \eqnref{eq:Ki}.
Therefore, the lattice model in \figref{fig:1fSMN} is a lattice realization of the $K$-matrix in \eqnref{eq:K}.

\end{document}